# Semantic knowledge guides innovation and drives cultural evolution

Anil Yaman*[1], Shen Tian*[2], Björn Lindström[2]

**1** Computer Science Department, Vrije Universiteit Amsterdam, Amsterdam, The Netherlands

**2** Division of Psychology, Department of Clinical Neuroscience, Karolinska Institutet, Stockholm, Sweden

* Equal contribution

Corresponding authors: a.yaman@vu.nl (AY) & bjorn.lindstrom@ki.se (BL)

**Abstract**

Cultural evolution allows ideas and technologies to accumulate across generations, reaching their most complex and open-ended form in humans. While social learning enables the transmission of such innovations, the cognitive processes that generate them remain poorly understood. Classical theories typically treat innovation as random variation, a simplification insufficient for explaining the complexity of human cultural evolution. We propose that semantic knowledge—the associations linking concepts to their properties and functions—guides human innovation and drives cumulative culture. To test this, we combined an agent-based model, which examines how semantic knowledge shapes cultural evolutionary dynamics, with a large-scale behavioral experiment (N = 1,243) testing its role in human innovation. Across both approaches, we found that semantic knowledge directed exploration toward meaningful solutions, enhanced innovation success, and enabled generalization from prior discoveries. Moreover, semantic knowledge interacted synergistically with social learning to amplify innovation and accelerate cumulative cultural change. In contrast, experimental participants lacking access to semantic knowledge performed no better than chance, even when social learning was possible, and relied on shallow exploration strategies for innovation. Together, these findings suggest that semantic knowledge is a key cognitive process underpinning human cumulative culture.

**Significance statement**

Human cumulative culture depends on our ability to innovate. While most research emphasizes how social learning spreads and preserves innovations, the cognitive processes that generate them remain poorly understood. Our study shows how semantic knowledge – our internal map of how concepts are related – guides innovation by directing exploration towards meaningful solutions. This cognitive process also works together with social learning to enhance collective innovation. Together, these findings identify a key cognitive foundation for human innovation, helping to explain its uniquely open-ended nature.

## Introduction

Cumulative cultural evolution, the process by which ideas and technologies accumulate across generations, underpins humanity's most significant achievements—from agriculture to artificial intelligence (1, 2). Cumulative cultural evolution is thought to be driven by two interacting processes: individual innovation, which generates and refines new ideas, and social learning, which allows these ideas to spread and accumulate (3). While the role of social learning is well established, the cognitive processes that make innovation possible remain poorly understood. Here, we investigate how semantic knowledge (4)—structured, generalizable associations between concepts and their properties and functions—guides human innovation by constraining exploration, and how it interacts with social learning to drive cultural evolution.

Although it may seem intuitive that semantic knowledge would facilitate innovation, this intuition is not directly captured in dominant theories of cultural evolution. Classical models of cumulative cultural evolution emphasize social learning and selective retention as drivers of cultural change, and often abstract innovation as random or weakly structured variation—an approach that has proven useful for modeling population-level dynamics (5). As a result, much theoretical and empirical work has focused on transmission processes and population-level selection, while the cognitive representations supporting individual innovation have remained less explicitly specified.

At the same time, these models allow for individual learning as a source of "guided variation", where socially acquired ideas are modified through individual experience, in cultural evolution (6). However, the cognitive mechanisms implementing such guided variation are typically represented as generic learning or inference processes (7). In this sense, existing theory formalizes the effects of individual learning on cultural change but leaves the cognitive substrate underspecified.

A complementary theoretical tradition, associated with the so-called Paris school, emphasizes how individual cognitive biases can shape cultural evolution even in the absence of strong selection (5, 8, 9). While this approach places cognition at the center of cultural change, it typically takes a functional approach that is not focused on the underlying cognitive mechanisms.

As a result, the role of culturally acquired semantic knowledge as a cognitive substrate shaping innovation has not been explicitly modeled or empirically tested. Whether such semantic representations causally structure innovation and cumulative cultural evolution, therefore, remains an open theoretical and empirical question.

Consistent with the view that innovation is guided by internal cognitive structure, empirical evidence increasingly suggests that cultural variation is not the product of unguided exploration alone (5): even in isolation, individuals innovate more effectively than random exploration algorithms (10). Indeed, several recent proposals suggest various types of reasoning abilities (11) as essential for innovation, including causal reasoning (12–14) and technological reasoning (15, 16). Causal reasoning enables individuals to predict the outcomes of their actions, diagnose failures, and foresee improvements in existing solutions (17), while technical reasoning, which applies causal reasoning to physical objects, helps individuals develop and refine tools and technologies (16, 18). While these reasoning abilities are likely to be important for innovation, they do not operate in a vacuum. Humans possess a rich base of semantic knowledge, which shapes what actions are considered and how problems are represented.

Semantic knowledge consists of structured, generalizable associations between concepts and objects, their properties, and typical functions, stored in semantic memory (19–21). Acquired through experience and language, semantic knowledge enables individuals to organize information, make analogies, and generalize to new situations. For example, knowing that sharp stones cut, fire heats, or plant fibers bind reflects semantic knowledge about typical object functions. In innovation, semantic knowledge supports intuitive understanding of object functions without requiring mechanistic insight (22). This distinguishes semantic knowledge from causal and technical reasoning capacities previously linked to innovation. While we focus on semantic knowledge about objects and their typical uses, as this domain is central to theory in cultural evolution (16), semantic knowledge also includes social domains (e.g., knowledge of social roles such as teachers and students (23), or emotion–situation associations such as loss causing sadness (24)). Crucially, semantic knowledge is itself a product of cultural evolution: it is socially transmitted, adapted to ecological and technological contexts shaped by prior generations' innovations, and refined across generations (25). As such, it represents not only an input to innovation but also an evolving output of cumulative cultural processes. Despite its foundational role in human cognition (4) and its significance in theories of human creativity (26–28) and tool use (29), its specific contribution to innovation and cultural evolution remains theoretically underarticulated and empirically untested.

To test the role of semantic knowledge in innovation and cumulative cultural evolution, we combined agent-based simulations and a large-scale behavioral experiment (N = 1,243) using a combinatorial innovation task designed to reflect core features of cultural evolution (10, 30). In this task, individuals attempt to innovate by combining different items (e.g., a sharp stone or a branch), either on their own or in groups, where social learning is possible. Innovation success results in new items (innovations, e.g., a stone axe), which can be used for future innovations. Such cumulative complexity is a hallmark of human cultural evolution (31). As in real-world innovation, the space of possible actions is vast, while only a small fraction result in successful discoveries. Efficient innovation, therefore, depends not only on retaining successful discoveries but also on generating plausible candidate actions in the first place. We hypothesize that semantic knowledge plays this guiding role by constraining exploration to actions that "make sense," thereby reshaping the effective search space on which cumulative cultural evolution operates.

We first used the agent-based simulation model to examine how semantic knowledge might enhance cultural evolution by guiding innovation. Unlike most previous models, which rely on simplistic agents without knowledge, our approach employs neural network–based representation learning to directly model semantic knowledge (32, 33). Semantic knowledge is not static: as individuals innovate, their internal knowledge representations change, and more successful individuals are more likely to transmit their knowledge to others. This reflects how semantic knowledge both drives and emerges from cultural evolution. Incorporating semantic knowledge changes cultural evolutionary dynamics: it focuses exploration on plausible actions and, in synergy with social learning, markedly enhances cumulative cultural evolution.

Next, we empirically tested the role of semantic knowledge in human innovation using the same combinatorial innovation task (10). We contrasted two experimental conditions: in the semantic condition, participants interacted with (depictions of) real-world items (e.g., stone, branch), allowing them to apply their prior semantic knowledge about how these objects typically function. In the non-semantic condition, the same items were represented by abstract symbols, removing access to real-world semantic knowledge (34). In both conditions, individuals either took the task individually or in groups, with the latter enabling real-time social learning. This design allows us to test whether semantic structure plays a causal role in innovation, or whether exploration, reasoning capacities, and social learning are sufficient to support cumulative discovery in its absence.

As predicted, we found that innovation success drastically decreased when participants lacked access to familiar semantic knowledge, and when social learning was possible. Consistent with the model, semantic knowledge and social learning synergistically enhanced innovation success, and semantic knowledge helped to narrow the action space, guiding participants towards more probable combinations. By analysing behavioral strategies, we found that participants used semantic similarity to generalize task-relevant knowledge and to combine semantically dissimilar items.

Together, our theoretical and empirical findings offer key insights into the cognitive foundations of cumulative cultural evolution, emphasizing how culturally evolved semantic knowledge structures innovation and, in turn, shape cultural evolution.

## Model overview

In the model, a population of individuals repeatedly attempts to innovate by combining items in an environment with an underlying combinatorial innovation task structure. Only specific combinations generate successful innovations; new items that can, in turn, serve as components for further innovations. Individuals initially possess six basic items and no knowledge of valid combinations.

Individuals vary in their capacities for semantic knowledge and social learning. These capabilities are controlled by parameters ranging from 0 to 1, where 0 indicates the absence of the ability, and 1 indicates maximal use. An individual with both parameters set to 0 relies entirely on random choices, lacking both semantic guidance and the ability to learn from others.

Each individual performs a fixed number of innovation attempts (10 in the base simulations). In each attempt, individuals either explore on their own or learn socially from the most successful individual in the population. During individual exploration, an individual selects one to three items from its current inventory and attempts new combinations, sometimes guided by its internal semantic model (see “Modeling semantic knowledge” below). Successful combinations produce new items, which are added to the individual’s inventory. Successful innovations also increase the individual’s score, with later and more complex discoveries yielding higher rewards. During social learning, an individual inspects the repertoire of the most successful individual (the one with the highest score) and copies a recipe (i.e., a specific item combination that leads to a successful innovation) it does not yet possess, but only if it can produce it with the existing items in its inventory.

After completing their innovation attempts, individuals update their internal semantic model based on all recipes they have acquired, regardless of whether they are obtained through individual discovery or social learning. The population then undergoes partial turnover through a Moran-style selection process: some individuals die and are replaced by offspring of more successful individuals, creating overlapping generations (35). Offspring inherit their parent's semantic model but begin with only the basic inventory of items. As a result, they must reconstruct specific items through individual innovation or social learning while benefiting from accumulated semantic knowledge. This assumption captures the idea that cultural inheritance primarily transmits generalizable knowledge rather than specific artifacts or procedural 'recipe' knowledge. Repeating this cycle over time allows the cultural repertoire and semantic knowledge to co-evolve.

We first compared populations with different social learning and semantic model capacities, and then allowed these capacities to compete within the same population. For all key assumptions and parameters, we conducted extensive robustness and sensitivity analyses (see "*Model sensitivity and robustness analyses"* below).

**Modeling semantic knowledge**

Individuals with semantic memory (in short, a "semantic model") can predict what combinations of items are likely to result in successful innovations (new items) (see Figure 1b & c). This memory is built upon a distributional semantic model, in which items are encoded as vectors in a high-dimensional space (36). In this space, the meaning of an item is defined by its functional context: its patterns of co-occurrence with other items in successful recipes (32, 37). Initially, all items are represented by vectors near the origin, reflecting the absence of functional relationships among them. As individuals discover or socially observe successful innovations, the model updates these representations through learning. Items that appear in similar functional contexts (e.g., red paint and blue paint, both of which can be combined with a log) acquire similar representations, causing functionally related items to cluster within a semantic similarity space (see Figure 1d). All items have vector representations, but only items that have appeared in successful innovations acquire meaningful structure. Items not yet involved in any successful combination retain uninformative representations and therefore do not contribute to the overall vector organization.

The semantic model supports similarity-based generalization by biasing exploration toward combinations involving semantically related items (see Methods for details), mirroring humans' similarity-based generalization (38, 39). For example, if combining red paint with logs is successful and blue paint has a similar representation, the model increases the probability of attempting the analogous combination (Figure 1d-e). Beyond the inductive bias provided by these learned representations, the model also included an explicit generalization mechanism (governed by $P_G$), in which individuals substitute an item of a successful recipe with a semantically similar item to innovate. In both ways, semantic similarity prioritizes exploration toward functionally plausible combinations.

**Model sensitivity and robustness analyses**

In the Supporting Information, we provide a comprehensive set of sensitivity and robustness analyses showing that our conclusions hold across variations in (i) the structure of the innovation task, (ii) the presence and mechanisms of vertical transmission, and (iii) the fidelity of vertical transmission. Additionally, we examined the impact of (iv) population size, (v) the probability of semantic model use, (vi) the mechanisms and probability of social learning, (vii) the probability of semantic recipe generalization, (viii) the number of innovation attempts per generation, (ix) the initial distribution of learning strategies in the population, (x) the cost of using the semantic model, and (xi) the neural network architecture (*SI: Figures S2–S14*).

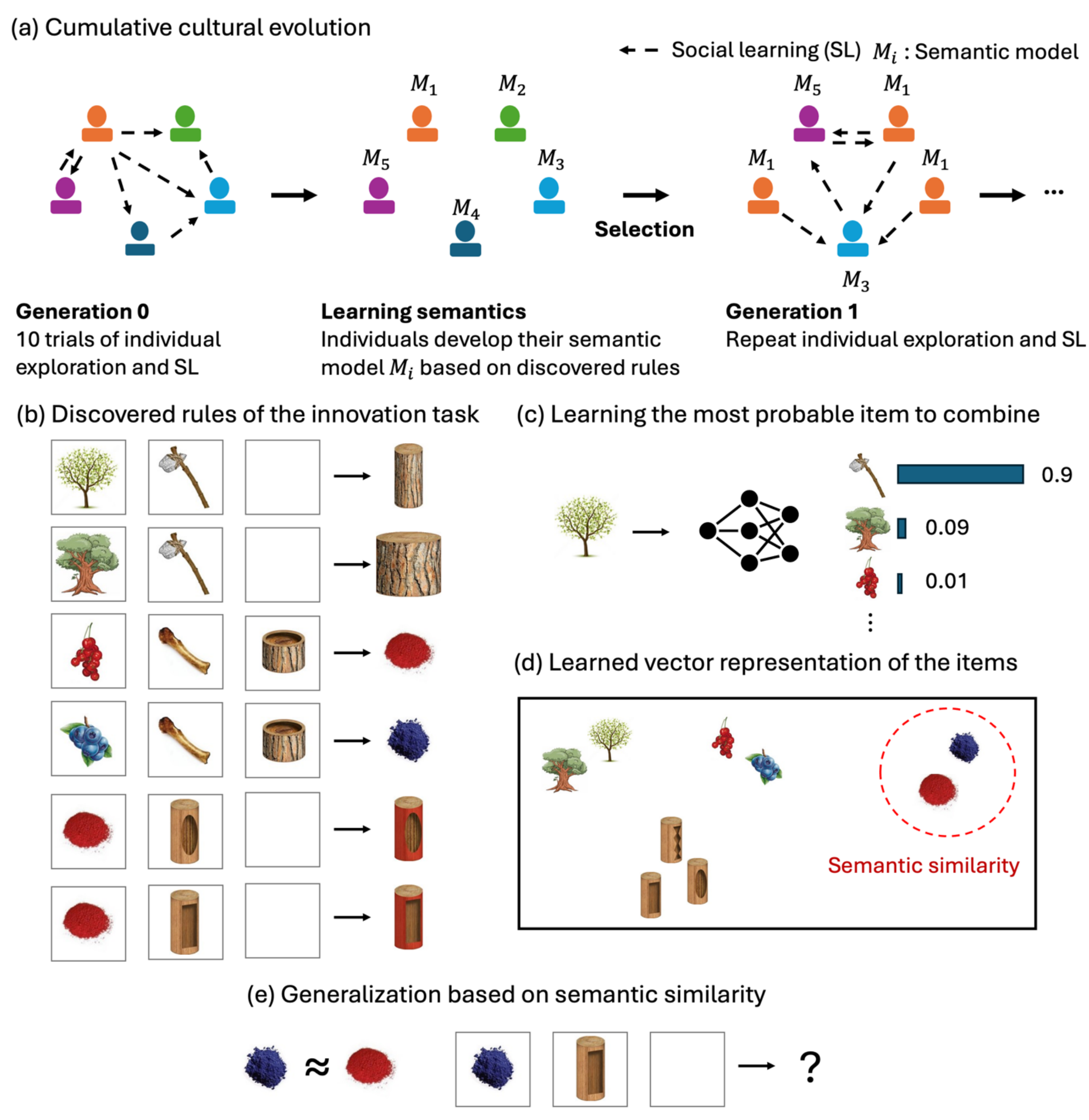

**Figure 1: The agent-based model of cumulative cultural evolution.** **(a)** Overview of the simulation. Individuals attempt to innovate by combining items and develop semantic knowledge through exposure to successful combinations (recipes). The initial population begins without knowledge of how items can be combined. Selection occurs each generation, with individuals more likely to reproduce if they are more successful innovators. Offspring inherit their parent's semantic knowledge but begin with the same item inventory as the initial population. **(b)** Examples of combinations resulting in successful innovations. Each row represents one recipe. **(c)** The semantic knowledge model enables individuals to predict which item combinations are most likely to succeed. **(d)** The model learns semantic representations of items, such that items with similar meanings have similar

internal representations. **(e)** These learned representations support similarity-based generalization, biasing future innovation attempts toward items semantically similar to those in past successful combinations. Images in b-d are taken from the task developed by Derex & Boyd (10).

## Results

### The role of semantic knowledge in cumulative cultural evolution

We first tested whether semantic knowledge enhances cumulative cultural evolution by comparing the growth of the cultural repertoire, defined as the number of different, unique items found by the population, in simulated populations with and without this capacity (Figure 2a). We also examined the role of success-biased social learning, which is well-established as crucial for cumulative cultural evolution (40).

As expected, social learning significantly increased the cultural repertoire (Figure 2a) by allowing innovations to spread through the population. However, cultural progress soon plateaued as the size of the innovation action space expanded. Crucially, semantic knowledge produced an effect comparable to social learning: populations endowed with semantic guidance accumulated substantially larger repertoires than those without it. However, these populations also plateaued because, without social learning, every individual must independently reconstruct every intermediate step (see *SI: Figure S3* for simulations delineating the role of vertical semantic transmission for these results). Similarly, individual random populations plateaued early, as cumulative rediscovery costs increased while the probability of identifying valid higher-level combinations decreased with the expanding action space (Figure 2a).

When both capacities were present, their effects were synergistic: populations combining semantic knowledge and social learning achieved roughly twice the repertoire of those relying solely on social learning. This synergy arises because social learning supplies additional successful recipes that serve as training data for the semantic model. The resulting refinement of semantic knowledge then improves individuals' ability to predict plausible new combinations, thereby fostering not only faithful reconstruction of existing cultural products but also the discovery of novel innovations. The result is an acceleration of cumulative cultural evolution. Increasing the probability of social learning up to an intermediate level further amplified this effect, but only in populations with semantic knowledge (*SI: Figure S8*). In contrast, synergy was absent for average individual innovation success (*SI: Figure S16*), where the effect of social learning also reflects direct copying of others' successful innovations, which does not require semantic knowledge. Thus, synergy emerges only at the population level, where shared learning updates semantic knowledge and generates a broader range of discoveries. These findings were robust across a range of model parameter values and alternative assumptions, including variations in task structure, the cost of semantic knowledge, transmission fidelity, and inheritance mechanisms (*SI: Figures S2-S14*).

We then ran evolutionary simulations in which semantic knowledge and social learning could evolve and compete directly within the same population (Figure 2b). Because the benefits of social learning often depend on its prevalence in a population (i.e., it is frequency-dependent (41–43)), it is critical to examine how it interacts with semantic knowledge over generations. Starting from a population of pure social learners without semantic knowledge, the type with both semantic knowledge and social learning rapidly became the dominant (Figure 2b), reinforcing our earlier conclusion that semantic knowledge strongly facilitates innovation and cumulative cultural evolution. This pattern was robust across different starting proportions of strategies (*SI: Figure S11*) and parameter values (*SI: Figure S15*).

Semantic knowledge emerged and became increasingly structured over generations (Figure 2c). To visualize this process, we represent the semantic models' embedding vectors as networks, where nodes correspond to items and edges reflect semantic similarity. Early in the simulation, individuals' knowledge networks were small and diffuse, reflecting limited associations between items. Over successive generations, these networks expanded and developed pronounced clustering, indicating the accumulation of organized semantic structure through cultural transmission. We next asked whether this culturally evolved structure resembled real-world human semantic representations. To test this, we compared the semantic representations produced by our model in the last generations with those generated by Sentence-BERT (44), a large language model trained on extensive text data (see *SI: Methods*). In both cases, each item (e.g., "stone", "branch") was represented as a vector in a high-dimensional space, where items with similar meanings were located closer together. We then measured how similar each item was to every other item in both models, producing two similarity matrices that we compared directly. The two were strongly correlated (Spearman $\rho = 0.65$, $p < .001$, see also *SI: Figure S18*), indicating that the semantic knowledge evolving in our model closely mirrors the structure of real-world human semantic knowledge.

Innovation requires exploring a vast space of possible actions. This is also true in our model: even when an individual has only ten items in their inventory, there are 285 unique possible combinations of items (combining 1–3 items at a time with replacement, ignoring order). In other words, constraining exploration to plausible combinations is key to successful innovation. To examine how semantic knowledge, together with social learning, constrains the action space, we took two approaches. First, we assessed the overall size of the action space by counting the number of unique actions as a function of inventory size during the final ten generations. As expected, individuals with semantic knowledge or social learning explored fewer unique combinations, avoiding options that did not make sense (Figure 2d). Second, we computed the Shannon entropy of action distributions within each inventory state. Higher entropy indicates more random exploration, while lower entropy reflects more focused, targeted search. As expected, entropy was markedly lower in populations with semantic knowledge (Figure 2e), indicating that individuals explored fewer, more targeted combinations than individuals without such knowledge. Entropy was also lower in populations with the capacity for social learning, consistent with the idea that social learning reduces individual exploration (45). These results confirm that both semantic knowledge and social learning restrict the action space, making innovation more efficient.

Overall, our results show that while social learning is a key driver of cumulative cultural evolution, semantic knowledge might play an equally critical role. When combined, these capacities enable populations to innovate and accumulate knowledge more effectively than when either capacity is present alone.

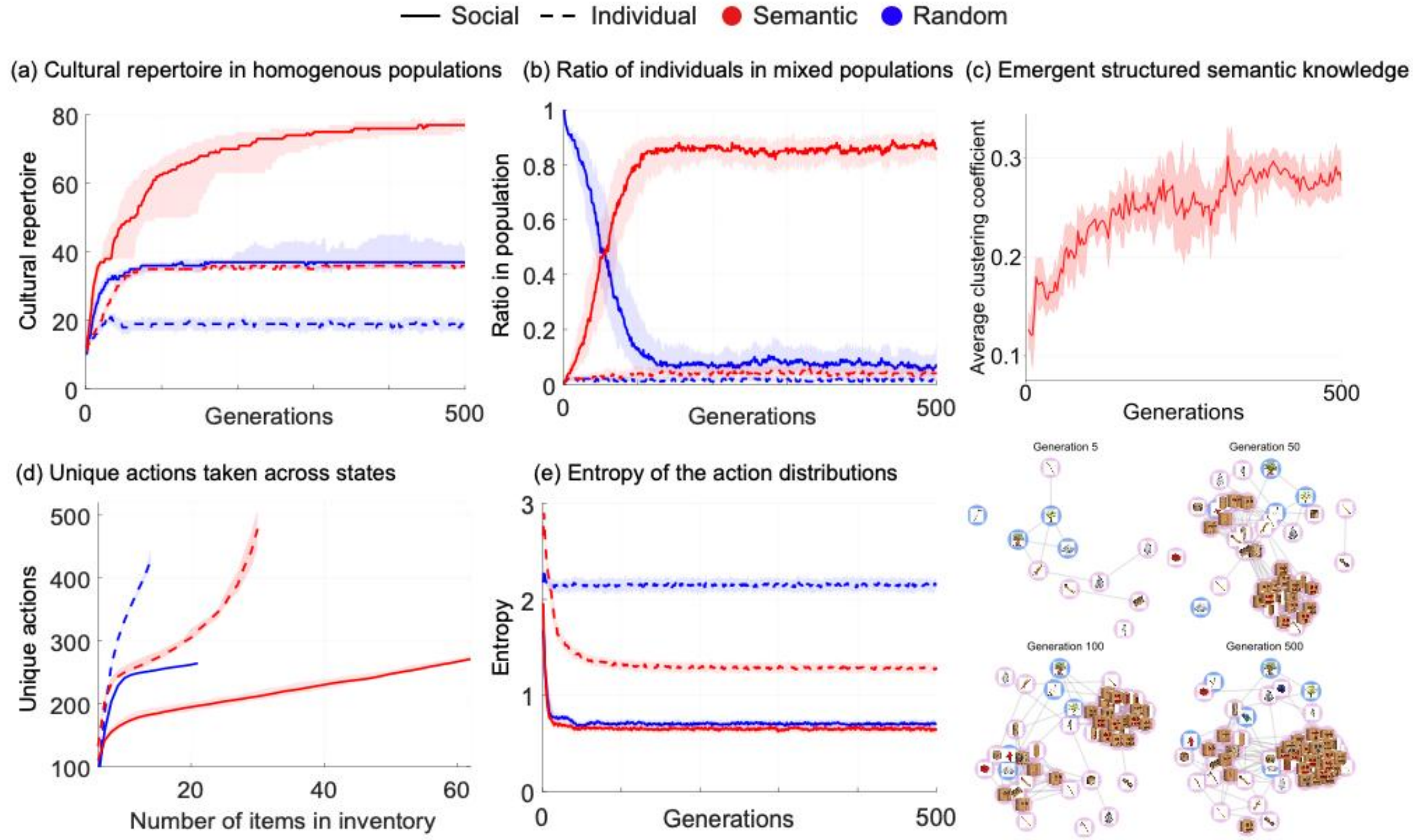

**Figure 2: Semantic knowledge drives cumulative cultural evolution. (a)** Populations with both semantic knowledge and social learning achieve a larger cultural repertoire (the total number of unique items in the population at a given time point) compared to populations lacking one or both capacities. **(b)** In mixed populations with four types of individuals, individuals with both semantic knowledge and social learning become dominant over time. The simulation started from a population of random social learning individuals without the capacity for semantic knowledge. **(c)** Semantic knowledge itself culturally evolves to become increasingly structured, as reflected in rising clustering coefficients (top; averaged across 20 individuals per generation from 5 independent simulation runs). Example semantic knowledge networks of the best-performing individual at four generations (bottom) illustrate the emergence of an organized structure. In these networks, the six basic items are represented by nodes with a blue background, and successful innovations are denoted by nodes with a plum background. **(d)** Individuals with semantic knowledge and social learning explore a smaller, more focused action space, taking fewer unique actions overall. **(e)** They also exhibit lower entropy in their action distributions (item selections) indicating that they selected similar items to combine guided by semantic knowledge and thus explored less randomly. Lines indicate the median of 96 independent simulation runs; shaded areas show the 1st and 3rd quartiles. Simulations used the following parameters: population size = 100, $P_{SL}$ = 0 (individual learning) vs $P_{SL}$ = 0.5 (social learning), and individual model parameters: $P_s$ = 0.9 and $P_g$ = 0.1 (see Methods). Similar patterns were observed across a range of parameter settings and model variations (see Figures S2-S14).

**Empirical evidence that semantic knowledge is crucial for human innovation**

Our theoretical model results suggest that semantic knowledge plays a key role in innovation and, consequently, human cultural evolution. To empirically test this hypothesis, we conducted an experimental study (N = 1,243) using the same innovation task as in our simulation model. Participants attempted to innovate by combining items from their inventory; each successful innovation increased their inventory size and final score, which was then converted into a monetary bonus.

To isolate the role of semantic knowledge, we compared two conditions: semantic and non-semantic (see Figure 3). In the semantic condition, items were represented by semantically meaningful images (e.g., stone, log), allowing participants to infer potentially successful combinations based on their prior semantic knowledge (Figure 3a). This is the standard version of the innovation task (10). In the non-semantic condition, the same items were instead represented by abstract symbols, preventing participants from drawing on prior semantic knowledge while keeping the underlying rules identical (Figure 3a). The images were matched for objective visual distinctiveness across conditions using multiple similarity metrics (see *SI: Figure S19*). As in the model, participants completed the task individually or in groups of six, allowing for real-time social learning (see Methods).

We assessed the effect of semantic knowledge on innovation using two complementary metrics: (1) the cultural repertoire, defined as the number of unique items discovered by each group (analogous to the population repertoire in the ABM), and (2) average innovation success, defined as the number of successful innovations discovered by each participant (as in the ABM).

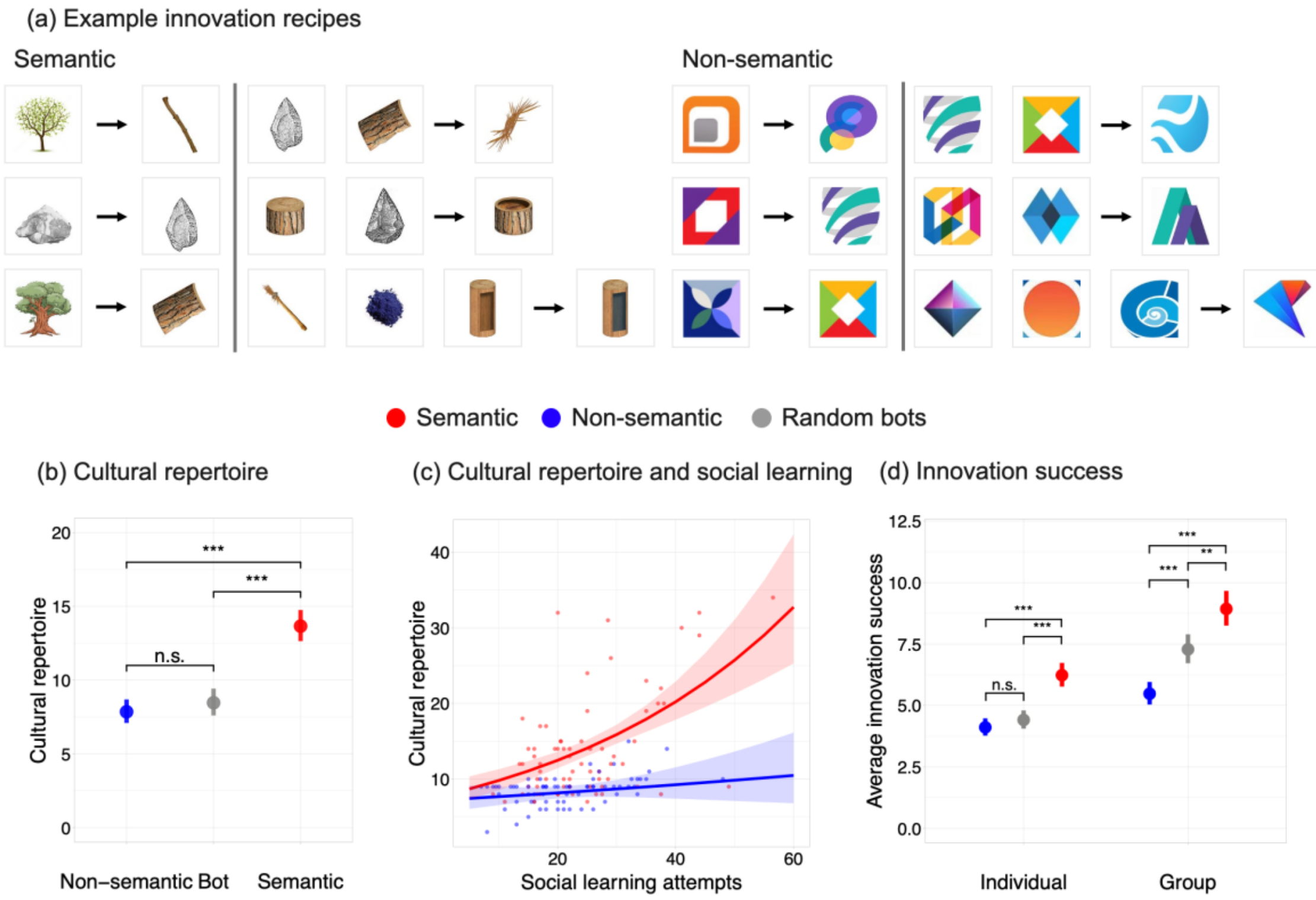

**Figure 3: Semantic knowledge guides human innovation. (a)** Experimental task. The semantic and non-semantic conditions had the same innovation recipes, but items were depicted with meaningful images in the semantic condition and abstract symbols in the non-semantic condition. **(b)** Participants generated larger cultural repertoires (total unique successful innovations per group) when semantic knowledge was available. Without semantic knowledge, human performance was comparable to random bots. **(c)** The difference between semantic and non-semantic conditions was moderated by the number of social learning attempts in the group, where more social learning resulted in larger differences between conditions. Bots had a fixed social learning probability ($P_{SL}$ = 1) and were therefore excluded from this analysis. **(d)** Individual participants also generated more successful innovations when semantic knowledge was available and social learning was possible. Dots and lines represent predictions from negative binomial regression models, and error bars and bands indicate model-derived 95% confidence intervals. *** p < .001, ** p < 0.01, n.s. p > 0.05.

Results confirmed our prediction that semantic knowledge is crucial for human innovation: groups in the semantic condition reached a larger unique cultural repertoire than those in the non-semantic condition (β = 0.55, SE = 0.06, z = 8.89, p < 0.001, Figure 3b). Consistent with the agent-based model, the impact of semantic knowledge on the cultural repertoire was amplified by social learning, with groups that engaged more in social learning benefiting disproportionately from semantic knowledge (interaction effect: β = 0.02, SE = 0.007, z = 2.90, p < 0.01, Figure 3c). This pattern mirrors the synergy observed in the simulations, where social learning provided new examples that expanded and refined knowledge, thereby accelerating innovation. In contrast, when semantic knowledge was unavailable, the size of participants' cultural repertoire was comparable to that of "bots" (see SI: Methods), which combine items randomly ($\beta = 0.07$, SE = 0.09, z = 0.92, p = 0.36, Figure 3b).

To further understand how semantic knowledge shaped human innovation, we next analyzed average innovation success at the participant level, directly comparing individual and group conditions. In the semantic condition, social learning in the group condition significantly enhanced average innovation success relative to the individual condition ($\beta = 0.36$, SE = 0.06, z = 6.41, $p < 0.001$), and individual humans outperformed random bots ($\beta = 0.34$, SE = 0.06, z = 5.97, $p < 0.001$). Both results replicate previous findings (see also *SI: Figure S20*) (10). As expected, innovation success was markedly reduced in the non-semantic condition ($\beta = -0.42$, SE = 0.06, z = -7.13, $p < 0.001$). Also consistent with the model, there was no significant interaction between semantic condition and group context (interaction effect: $\beta = 0.07$, SE = 0.08, z = 0.87, p = 0.38, Figure 3d). This reflects the fact that social learning can boost individual success through direct copying, even without semantic knowledge, which improves performance but does not expand the group's cultural repertoire.

Notably, without semantic knowledge, humans performed on par with random bots when innovating alone ($\beta = -0.07$, SE = 0.06, z = -1.17, p = 0.85), and benefited less from social learning in groups than bots (interaction effect: $\beta = 0.21$, SE = 0.08, z = 2.52, $p < 0.05$). This indicates that when deprived of semantic knowledge, humans relied on alternative strategies (see Section 3.3) that less efficiently supported innovation success. Together, these findings provide strong empirical support for our hypothesis that semantic knowledge guides human innovation.

**Semantic knowledge narrows the action space and guides innovation**

To better understand the role of semantic knowledge in human innovation, we examined whether and how semantic knowledge (i) narrows the action space, (ii) facilitates generalization, and (iii) guides exploration.

Our model results suggest that semantic knowledge helps individuals to focus their innovation attempts on combinations that "make sense." This should manifest in two ways. First, participants in the semantic condition should avoid meaningless combinations, leading to fewer unique combination actions. As expected, larger inventories increased the number of unique combinations (linear mixed model, controlling for non-independence within individuals and states: $\beta = 11.40$, SE = 0.34, $t = 33.35$, $p < 0.001$), but this effect was as attenuated in the semantic condition (interaction: $\beta = -16.70$, SE = 1.09, $t = -15.30$, $p < 0.001$), see Figure 4a. Consistent with our simulation results, the action space was also smaller in the group condition where social learning was available (interaction: $\beta = -22.3$, SE = 0.87, $z = -25.62$, $p < 0.001$). Second, if semantic knowledge directs participants toward meaningful combinations, they should engage in less random exploration. We quantified this by measuring the entropy of the action distribution, where higher entropy reflects more random exploration. As in the simulation, entropy was lower in the semantic condition than in the non-semantic condition (linear mixed model: $\beta = -0.70$, SE = 0.09, $t = -7.93$, $p < 0.001$), indicating that participants explored fewer, more targeted combinations (see Figure 4b). Similarly, entropy was also lower in the group condition than in the individual condition ($\beta = -0.58$, SE = 0.07, $t = -7.86$, $p < 0.001$), confirming that both semantic knowledge and social learning constrain the action space and enhance innovation efficiency.

Our agent-based model included semantic generalization, which allowed individuals to predict successful innovations based on their similarity to previous experiences (Figure 1e). This type of similarity-based generalization is fundamental to the underlying neural network model. We tested whether human participants also generalize based on semantic similarity, even if it is not guaranteed to match the hidden structure of the task. If so, they should preferentially try item combinations that were semantically similar to those used in previous successful innovations, regardless of knowing whether they are successful or not. We measured semantic similarity between the items used in each successful innovation and those in the next attempt, and compared it to the similarity following unsuccessful attempts. Consecutive attempts were more semantically similar in the semantic condition than in the non-semantic condition ($\beta = 0.03$, SE = 0.003, $z = 8.41$, $p < 0.001$) or random bot baselines ($\beta = 0.04$, SE = 0.004, $z = 10.87$, $p < 0.001$; Figure 4c). Crucially, this effect occurred only after successful attempts (interaction with semantic condition: $\beta = -0.1$, SE = 0.02, $t = -4.38$, $p < 0.001$; interaction with non-semantic condition: $\beta = -0.04$, SE = 0.03, $t = -1.40$, $p = 0.16$), indicating that participants generalized selectively based on success (39). This pattern cannot be explained by a systematic search strategy (e.g., sequentially combining one item with all others, independent of success), supporting the conclusion that participants use semantic generalization to guide exploration.

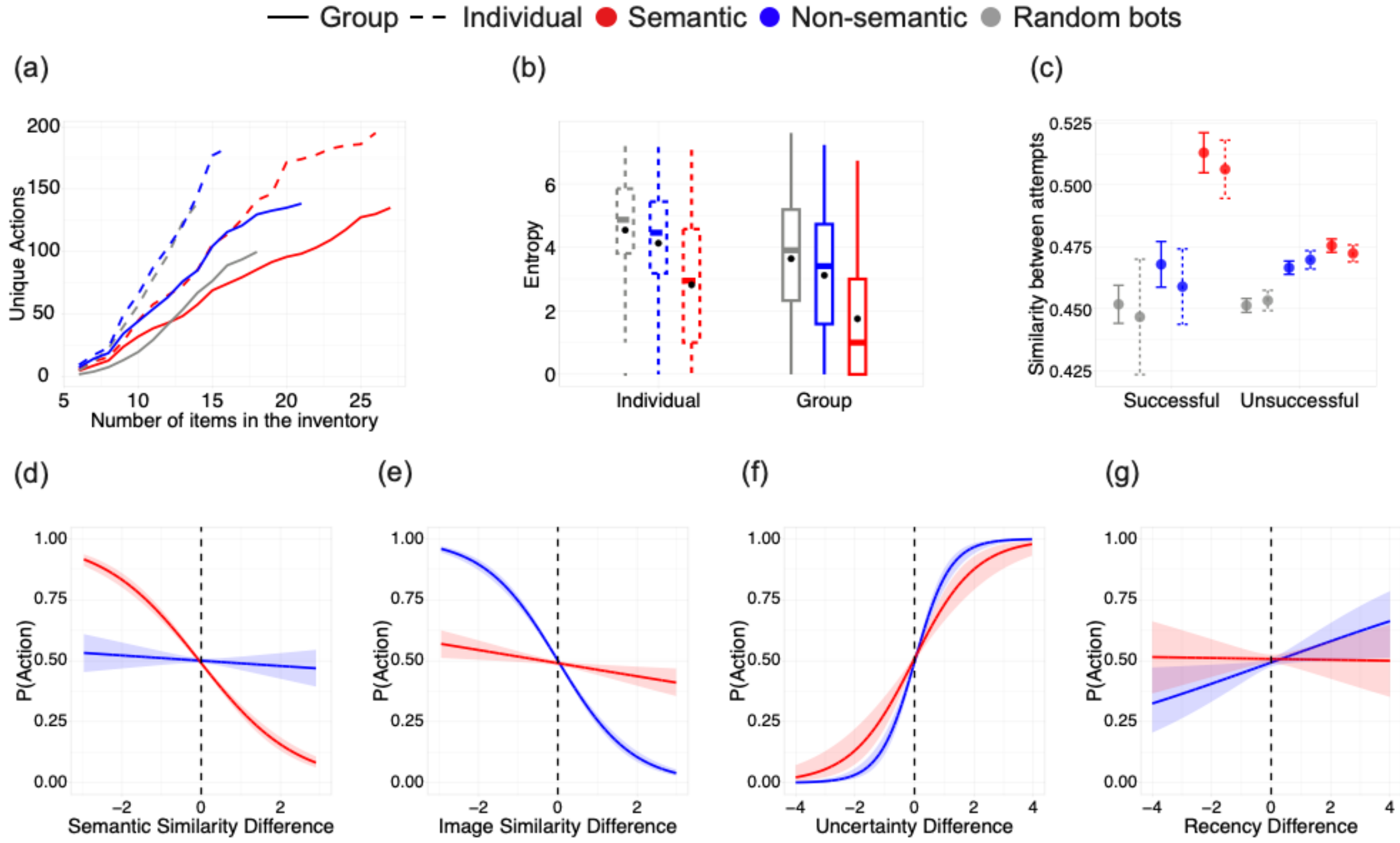

**Figure 4**: **Semantic knowledge reduces the action space and shapes the behavioral strategies used for innovation.** **(a)** Participants made fewer unique item combinations in the semantic and group conditions across inventories (states). **(b)** Lower entropy in these conditions indicates less random exploration compared to the non-semantic condition and random bots. Boxes show the first to third quartiles of the distribution, whiskers show the range of the data, and black dots indicate condition means. **(c)** Participants in the semantic condition exhibited semantic generalization, using items similar to those in previous successful innovations. Dots represent predictions from linear mixed regression models, and error bars indicate model-derived 95% confidence intervals. **(d)** Participants relied on semantic dissimilarity in the semantic condition, whereas **(e)** they relied more on perceptual image dissimilarity in the non-semantic condition. The y-axes show the difference between the feature value of the chosen item combination relative to alternative combinations. **(f-g)** Participants in the non-semantic condition relied more on uncertainty **(f)** and recency **(g)** than participants in the semantic condition. The analysis of behavioral strategies combined data from the individual and group settings. Confidence bands indicate 95% confidence intervals. Differences in features were calculated by comparing participants' actual actions with alternative actions.

Next, we investigated how participants leveraged semantic knowledge when selecting item combinations. One possibility is that participants combine semantically similar items, reflecting how more similar words tend to become activated in semantic memory during semantic search (46). On the other hand, prior work suggests that innovation often requires combining distant concepts (47, 48). For example, creating an axe requires combining semantically distinct components: a cutting tool (e.g., *stone*) and a handle (e.g., *stick*).

To test these alternative hypotheses, we used a generalized linear mixed model (GLMM) to model action selection. Because the action space grows exponentially with inventory size, we reduced the problem to logistic regression, comparing participants' actual item combinations to randomly sampled combinations from the same inventory state (49) (see Methods for details). This allowed us to estimate how features such as semantic similarity influenced the probability of selecting a given combination.

Semantic similarity between items was quantified using a pre-trained sentence embedding model (Sentence-BERT (44)) that converted item labels (e.g, "small stone") into vector representations, from which we computed the pairwise cosine similarity between all vectors (see *SI: Methods*). We also included perceptual similarity, computed via image structural similarity and color distribution similarity metrics (see *SI: Methods*), to account for the possibility that participants in the non-semantic condition relied more on low-level visual cues. Because abstract images in the non-semantic condition lacked inherent meaning, we used semantic similarity values derived from the corresponding real-world items for both conditions. Naturally, there should be no effect of this predictor in the non-semantic condition. Analyses included only unsuccessful innovation attempts to control for task structure effects, as successful combinations were inherently more semantically dissimilar (pairwise $r_{successful\ combinations}$ = 0.19 < $r_{all\ combinations}$ = 0.61). For robustness, we also conducted an alternative set of analyses based on a behavioral representational similarity analysis (RSA) approach (50), which produced equivalent results (see *SI: Figure S21*).

Participants preferred combining semantically dissimilar items, but only when semantic knowledge was available (interaction effect: $\chi^2(1) = 258$, $\beta = -0.78$, SE = 0.08, $z = -10.3$, $p < 0.001$; Figure 4d). This serves both as a sanity check for the analytical approach and as evidence that participants use semantic dissimilarity to guide innovation. Furthermore, participants also tended to combine items that are more structurally dissimilar ($\chi^2(1) = 487$, $\beta = -1.00$, SE = 0.06, $z = -17.3$, $p < 0.001$), an effect that was stronger in the non-semantic condition (interaction effect: $\chi^2(1) = 556$, $\beta = 0.88$, SE = 0.07, $z = 12.5$, $p < 0.001$) (Figure 4e). This suggests that when semantic knowledge is available, it serves as the primary guide for innovation, while in its absence, people shift to alternative strategies, such as visual similarity.

Next, we extended our model to include additional strategies that might shape human innovation, such as reward history, prior success, inventory position, number of items per combination, uncertainty (51), recency, and empowerment (favoring items that could unlock future successful innovations (49)), see *SI: Methods* for definitions and *SI: Figure S23* for full statistics and their distributions.

Semantic similarity remained a strong predictor of innovation attempts (interaction effect: $\chi^2(1)$ = 147, β = -0.81, SE = 0.11, z = -7.42, p < 0.001), confirming the robustness of the preceding analysis. Similarly, in the non-semantic condition, the influence of image similarity remained significant (main effect of structural similarity: $\chi^2(1)$ = 1.55, p > 0.05, interaction with semantic condition: $\chi^2(1)$ = 75.8, β = 0.49, SE = 0.11, z = 4.53, p < 0.001; main effect of color similarity: $\chi^2(1)$ = 53.6, p < 0.001, interaction with semantic condition: $\chi^2(1)$ = 84.1, β = 0.67, SE = 0.10, z = 6.45, p < 0.001). Additionally, we found clear evidence for several other strategies (see *SI: Figure S23b*). Most notably, participants in both conditions preferred more uncertain options ($\chi^2(1)$ = 355, p < 0.001), indicative of directed exploration (51) (Figure 4c, see *SI: Methods* for statistical details). However, this effect was significantly stronger in the non-semantic condition (interaction uncertainty: $\chi^2(1)$ = 17.7, p < 0.001, see Figure 4f) accompanied with a preference for items that had been recently used (interaction recency: $\chi^2(1)$ = 14.5, p < 0.01, see Figure 4g), indicating that in the absence of semantic information, participants relied more heavily on heuristic strategies to guide their exploration.

Together, the results suggest that semantic knowledge is central to innovation, guiding exploration toward meaningful combinations and enabling generalization from prior successes. When semantic knowledge is unavailable, participants instead rely on other cues, including perceptual similarity, uncertainty, and recency.

**The role of success-biased social learning in human innovation**

Consistent with previous research, both our simulation model and empirical results show that social learning supports innovation success. To better understand this, we examined *when* participants used social learning and from *whom* they learned.

Prior studies suggest that individuals are more likely to copy others when uncertain or dissatisfied (52, 53). To test this, we first analyzed whether the number of unsuccessful innovation attempts at time *t* would predict participants' tendency to observe others' successful innovations at *t+1,* using a negative binomial mixed effects model. Participants were more likely to observe others at *t+1* after experiencing more failures at *t* (β = 0.02, SE = 0.007, z = 3.36, p < 0.001), with no moderating effect of condition (interaction effect: β = 0.002, SE = 0.01, z = 0.22, p = 0.83), see Figure 5a. They were also more likely to observe others when their score was below the group average (β = 0.02, SE = 0.002, z = 7.31, p < 0.001). While overall rates of social learning did not differ between the semantic and non-semantic conditions (β = 0.05, SE = 0.13, z = 0.36, p = 0.72; Figure 5b), dissatisfaction had a weaker effect on social learning when semantic information was available (interaction: $\beta = -0.01$, $SE = 0.003$, $z = -4.39$, $p < 0.001$; Figure 5b). This suggests that semantic knowledge alters the motivation for seeking social information. In the absence of semantic knowledge, participants relied more on a copy-when-dissatisfied strategy, using social learning primarily in response to failed exploration. In contrast, when semantic knowledge was available, participants could guide search internally, reducing reliance on immediate success or failure cues (54).

To test whether participants exhibited success-biased social learning as in our model, we examined whether they preferentially learned from higher-scoring group members. Participants overwhelmingly chose to observe the top-ranked individual (estimated probability: 61% and 65% in semantic and non-semantic conditions, respectively; see *SI: Figure S24c*), and higher scores predicted being observed ($\beta = 0.47$, $SE = 0.02$, $z = 19.86$, $p < 0.001$). This bias was slightly weaker in the semantic condition (interaction: $\beta = -0.10$, $SE = 0.03$, $z = -3.38$, $p < 0.001$; Figure 5d), suggesting again that participants in this condition were more certain of their actions.

Together, these results indicate that participants relied on both personal outcomes and social cues to regulate social learning. Moreover, semantic knowledge appears to shift social learning from a reactive response to dissatisfaction toward a more strategic search for social information that complements individual exploration.

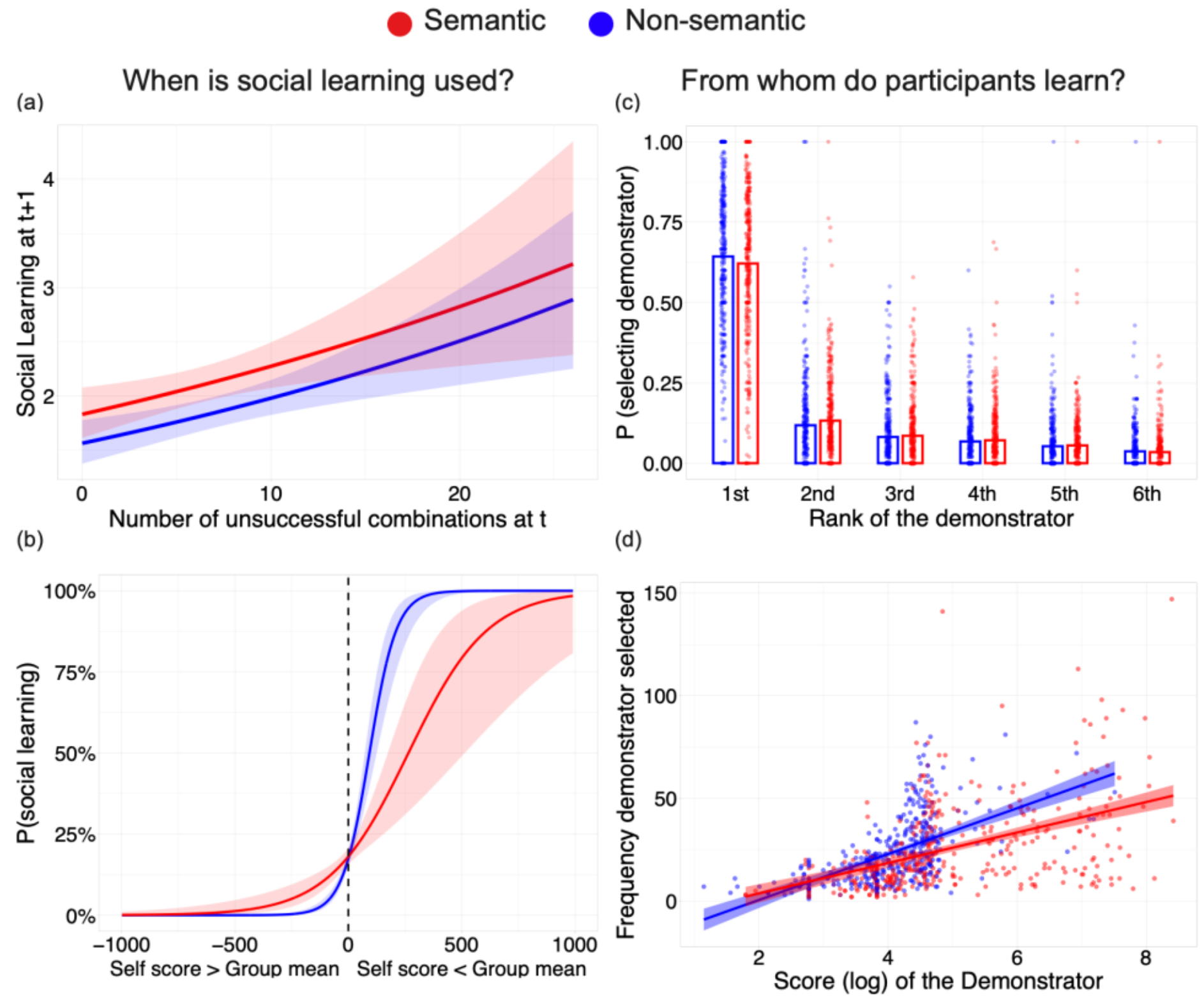

**Figure 5: Participants use success-biased social learning when dissatisfied. (a)** Participants increased social learning after experiencing more prior failures. **(b)** Social learning was guided by the participants' own performance and the success of the group members. Participants were more likely to learn socially when their own scores were below the average score of the group. This was particularly pronounced in the non-semantic conditions. **(c)** Participants in both semantic and non-semantic conditions preferentially learned from the most successful member in the group. **(d)** Participants with higher scores were more likely to be selected as demonstrators.

## Discussion

Here, we provide evidence for the crucial role of semantic knowledge in innovation and cultural evolution. While semantic knowledge has long been recognized as central to human cognition, our findings show its specific importance in shaping innovation and, in turn, cumulative cultural evolution. Using agent-based modeling and a large behavioral experiment, we show that semantic knowledge enhances innovation success by constraining the vast space of possible actions. In this way, it functions similarly to social learning, but through internalized, culture evolved knowledge rather than external social cues. Together, these capacities provide complementary scaffolds for cumulative cultural evolution. Crucially, our results show that semantic knowledge cannot be reduced to general exploration, reasoning ability, or social transmission. When semantic structure was removed, participants, despite intact motivation, feedback, and access to social information, failed to outperform randomly exploring bots. In other words, semantic knowledge is not merely facilitative but plays a constitutive role in shaping the innovation process itself.

A key insight from both the simulations and behavioral data is that semantic knowledge and social learning can interact synergistically. Each process independently enhances innovation: semantic knowledge narrows exploration to meaningful possibilities, and social learning diffuses successful innovations. Their combination, however, can produce more than additive gains. This synergy arises because observing others' innovations provides information that improves individuals' semantic models, which in turn guides their own exploration. These findings suggest that cumulative cultural evolution is particularly effective when semantic knowledge and social learning act synergistically.

Unlike earlier knowledge-free models of cultural evolution, our agent–based modeling approach explicitly represents semantic knowledge using distributional methods inspired by natural language processing (19, 37, 46). Individuals learn that items used in similar contexts share similar functions, allowing them to infer likely successful innovations and transmit this knowledge to the next generation. Sensitivity and robustness analyses confirmed that these effects hold across a range of model parameters and alternative assumptions (see *SI: Results* and *Figure S2-S16*). The resulting culturally evolved knowledge closely resembles human semantic knowledge, as indicated by the strong correlation between our model's semantic representations and those derived from large-scale text corpora (44). Although the model assumed direct (but noisy, *SI: Figure S6*) inheritance of knowledge, future work could explore alternative mechanisms, such as language-mediated communication, which could facilitate abstraction (55) and use of analogy (11, 56) in response to different transmission fidelities. Finally, our approach focused on semantic knowledge about objects and their functions, a core but not exhaustive aspect of human semantic memory. Other domains, such as social, emotional, or linguistic knowledge, may scaffold different forms of innovation and cultural accumulation (23, 24). Future work could extend our framework to these content domains.

Consistent with the theoretical model, our behavioral data show that without access to familiar semantic knowledge, participants performed markedly worse, even when social learning was available. Analysing the behavioral strategies underlying innovation, we found that in the absence of semantic knowledge, people relied more on shallow exploration strategies, such as recency, low-level perceptual features, and uncertainty-driven exploration (51). These strategies did not compensate for the lack of semantic knowledge, as individuals in the non-semantic condition performed on par with random bots. This highlights a critical implication: without structured semantic priors, human innovation reduces to near-random search. This suggests that the efficiency of human innovation does not primarily stem from general intelligence or reasoning alone, but from culturally accumulated semantic structure.

When semantic knowledge was available, participants favored semantically dissimilar item combinations, consistent with real-world innovation patterns that include both short-range, within-domain and long-range, cross-domain combinations (57). In contrast to some earlier studies, we found no clear evidence of empowerment-based exploration (49). This was likely due to the task structure (see *SI: Methods*) or the direct reward feedback. Nonetheless, semantic contrast—favoring dissimilar items—may serve as a mechanistic proxy for the "sense of success" component of earlier empowerment definitions (49), reflecting the selection of items likely to yield novel outcomes.

In both the agent-based model and the experiment, semantic knowledge reduced the space of actions that individuals explored. While random bots explore this space uniformly, semantic knowledge guides innovation attempts toward plausible, meaningful actions. By constraining exploration, semantic knowledge does not simply reduce the number of actions considered; it directs search toward regions of the innovation space where viable combinations are more densely connected. This is likely to become decisive when innovation spaces are large and sparse, conditions characteristic of real-world cultural evolution. Social learning can produce a similar focusing effect but also lead to maladaptive herding in suboptimal regions of the action space (56, 57), as seen in financial market bubbles (58) and panic buying during public health crises (58). Because semantic knowledge is shaped by broad, generalizable environmental regularities, it should typically be adaptive (59). However, strong semantic priors can also act as a double-edged sword by biasing exploration toward common-sense solutions and potentially overlooking counterintuitive or 'unreasonable' discoveries (60). Indeed, a hallmark of highly creative individuals might be the ability to repurpose familiar objects in atypical ways (61). Directly modeling environments in which semantic priors hinder innovation represents an important direction for future research.

More broadly, these results highlight parallels between innovation and research on creativity and associative thinking. Creativity is often conceptualized as a two-stage process: an associative search through semantic memory networks followed by an evaluation of candidate ideas based on novelty and usefulness (26). It is plausible that innovation reflects a similar semantic search mechanism (62). A goal of future research should be to uncover the computational mechanisms underlying semantic innovation by formulating and testing mechanistic process models.

Our results also have possible implications for human-AI collaboration. Specifically, we found that a simple, random exploration strategy performed as efficiently as human participants in the non-semantic conditions, and even outperformed them when social learning was available. This suggests that human innovation could be enhanced not only by sophisticated AI, such as large language models (63), but also by simpler, interpretable artificial agents (64), capable of exploring large action spaces to uncover unconventional solutions that humans might overlook.

There are several limitations worth noting. First, while the innovation task captured core features of real-world innovation, it was computerized. It did not involve interaction with physical objects, limiting the engagement of cognitive processes involved in technical reasoning (18). Second, the task simulates a closed-world innovation process, with a fixed set of items and predetermined combination recipes. However, because the full structure of the task is not revealed, participants must still engage in exploration to identify what combinations are possible.

More generally, the task is simpler and more causally transparent than many historically important real-world innovation problems, such as developing complex tools or discovering causally opaque processes (17). In such domains, selective cultural processes may play an especially important role in accumulating and stabilizing partial solutions beyond the reach of individual inference. Our results suggest that semantic knowledge nevertheless provides an important cognitive resource for structuring exploration, but its interaction with selective cultural dynamics in more complex and opaque problem spaces remains an important direction for future research.

Third, future research should also consider how people learn and update their semantic knowledge over time, especially in situations where existing semantic knowledge is unavailable or unhelpful. In such cases, participants may rely more on the model-free, trial-and-error learning. This raises questions about how people arbitrate between offline semantic models and online, experience-driven learning (65). Computational models that capture these dynamics could provide deeper insights into the cognitive mechanisms supporting innovation.

In conclusion, our findings provide a unique perspective on the cognitive foundations of innovation and cultural evolution. While traditional models of cumulative cultural evolution have emphasized social processes, such as social learning (1), network structure (66), and group diversity (67), they have often overlooked the role of human cognitive capacities in guiding innovation. Our results extend these accounts by demonstrating that semantic knowledge plays a crucial role in guiding innovation and, in extension, driving cumulative cultural evolution. These insights underscore the need to integrate cognitive mechanisms into theories of cultural evolution and offer future directions for understanding how structured knowledge and social processes together shape the accumulation of culture.

## Methods

### The cumulative cultural evolution model

We here outline the key components of the model. Please see *Figure S1* for an overview of the model, *SI: Methods* and *SI: Table 1-2* for additional details, and *SI: Algorithms 1-2* for pseudocode. An overview of sensitivity and robustness analyses is provided in *SI: Figure S2-16*.

**Population structure.** Our CCE model simulates populations of individuals evolving under a Moran-style selection process. Individuals perform a fixed number of innovation attempts (10 in the main simulations). After these attempts, selection occurs: individuals with higher scores are more likely to reproduce and transmit their semantic knowledge. Over time, older individuals are probabilistically replaced by offspring, generating an age-structured population.

**Individual model.** Individuals choose between individual and social learning based on a probability parameter, $P_{SL}$. When engaging in individual learning, the innovation attempt (item combination) is generated either randomly or guided by the semantic model, with probability $P_S$. If the semantic model is used, individuals select one item at random and then use the model to predict which other items are most likely to combine successfully with it. Individuals may employ an explicit generalization strategy, leveraging a previously successful combination by substituting one item for another with the smallest semantic distance, with probability $P_G$. During social learning, individuals observe the most successful individual in the population and copy a successful recipe only if they already possess all prerequisite component items. Thus, social learning never allows individuals to bypass intermediate steps or acquire innovations they could not independently produce from their current inventory.

**Learning semantic knowledge.** We modeled semantic knowledge using a skip-gram word2vec algorithm implemented as a feedforward artificial neural network (ANN) with one hidden layer (32). In this framework, items are represented as vectors in high-dimensional space. The network was trained to predict which items are likely to combine successfully (see Figure 1c), allowing items appearing in similar successful combinations to acquire similar representations. Because recipes could involve different numbers of items (1-3), we introduced a technical *null item* token. For recipes with fewer than three components, the null item served as padding, enabling all combinations to be encoded in a common format. Recipes with only one item represent processing of that item (e.g., tree -> stick). Successful innovations provided co-occurrence information that shaped the geometry of the semantic space. To maintain a globally consistent representation as new items and recipes were discovered, the semantic model was retrained periodically rather than updated after every innovation attempt. Retraining used the individual's entire set of successful recipes acquired through both individual exploration and social learning, allowing the vector space to reorganize as knowledge accumulated. We examined robustness across two key hyperparameters: the hidden layer width and embedding dimensionality. Networks with 16 or 32 neurons performed similarly well; we report results using 16 dimensions to balance accuracy and computational efficiency.

## Experimental study

**Participants.** A total of 1,243 participants (M = 35.6 years, SD = 12.3; 622 women) were recruited via Prolific Academic. Informed consent was obtained from all participants, and ethical approval was granted by the ethics board of Vrije Universiteit Amsterdam.

**Treatments:** Participants were randomly assigned to one of four conditions: individual non-semantic (N = 211), individual semantic (N = 210), non-semantic group (N = 408), and semantic group (N = 414). Group-condition participants were organized into groups of six.

**Task design and experimental procedure.** We adapted the innovation task from Derex & Boyd (10), simplifying its rules for online administration (~10 minutes) to reduce attrition. Participants began with 6 basic items and could discover up to 178 additional items through exploration. Each discovery contributed to a cumulative score used to calculate a performance-based bonus. On average, participants earned £2 for participation plus £1.60 in bonuses (range £0–£5.27). Before the task, participants received condition-specific instructions and examples. The experiment was run on a custom Python-based web platform (see *SI: Figure S17*).

## Human behavioral analysis

**Cultural repertoire.** We analyzed the number of unique items found by each group using negative binomial models, with the median number of social learning attempts in the group, the median number of completed innovation trials, and the number of players (to control for attrition) in the group as predictors.

**Action space.** For each unique inventory state, we counted the number of unique combination attempts and normalized it by the number of items available in that state. Counts were cumulative, meaning later states included actions from earlier states. We used Shannon entropy (68) to measure the uncertainty of participants' action distributions given a specific inventory state as: $H(x) = -\sum_{x \in X} p(x) log p(x)$, where $x \in X$ is the combination of items tried in that state. Higher entropy indicates more exploratory, less structured behavior.

**Number of innovations.** To compare innovation success across conditions, we fit a negative binomial mixed-effects model using the *glmmTMB* package in R (69), with nested random effects of participant ID and group ID. Pairwise post-hoc comparisons tested specific differences between conditions.

**Behavioral strategies**. We simulated datasets comprising participants' actual combinatorial behaviors and 'alternative behaviors' randomly sampled from the complete space of possible item combinations available at each inventory state. For each innovation attempt, various features were computed for both 'actual actions' and 'alternative actions' (see *SI: Methods* for feature definitions). The difference between these features, along with their interactions with the two experimental conditions (semantic vs non-semantic and individual vs group), was entered into a generalized linear mixed-effects model (GLMM) with a binomial error distribution (i.e., logistic regression) using the lme4 package (70). Attempts involving only a single item were excluded, since similarity features cannot be computed for them. The model included random intercepts for each participant and group, and by-participant random slopes for all fixed effects. Two models were tested: (1) including only semantic and perceptual similarity features, and (2) a comprehensive model with all behavioral features (see *SI: Methods*). To ensure robustness, Model 1 was fit 100 times with different sets of alternative actions, and Model 2 was fit 50 times (due to the high computational cost). The resulting distributions of estimates (see *SI: Figure S22-S23*) were averaged for the main text results.

**Semantic generalization.** To assess generalization, we measured semantic similarity between items used in each successful innovation and those in the immediately following attempt. Greater reuse of semantically similar items after success was taken as evidence of semantic generalization. We compared this effect following successful vs. unsuccessful attempts using linear mixed models with participant-level random slopes for all fixed effects.

**Social learning.** Social learning involves two steps: first, selecting a demonstrator and inspecting their inventory, and second, reviewing the recipe of a specific item. Events were logged whenever participants inspected a demonstrator's inventory. We recorded both the inspected items and the participant's current score. To analyze timing, we examined two predictors: (1) personal exploration history, quantified as the number of failures in the preceding minute (with experiments divided into 10 one-minute windows), and (2) performance gap, defined as the score difference between the participant and other group members. The likelihood of social learning was modeled using a negative binomial mixed-effects model. To analyze demonstrator choice, we tested the association between demonstrator scores and frequency of being observed. For ordinal analyses of demonstrator rank, we used the *clmm* function from the *ordinal* package in R (71).

**Data availability.** Data are available at the Open Science Framework at https://osf.io/m642a/ (72).

**Code availability.** Model and analysis code will be available at the Open Science Framework at https://osf.io/m642a/ (72).

## Acknowledgements.

We thank Alexandre Bluet, Lucas Molleman, Andreas Olsson, and David Schultner for their invaluable comments on an earlier draft of this paper. We thank Andrzej Szczepura for contributing to the development of the experimental task. This work received funding from the Computer Science Department, Vrije Universiteit Amsterdam, to AY, from a Wallenberg Academy Fellow grant from the Knut and Alice Wallenberg Foundation (KAW 2021.0148) to BL, and from a Starting Grant (SOLAR ERC–2021–STG–101042529) from the European Research Council, to BL. The funders had no role in study design, data collection and analysis, decision to publish, or preparation of the manuscript.

Supporting Information for

# Semantic knowledge guides innovation and drives cultural evolution

Anil Yaman*[1], Shen Tian*[2], Björn Lindström[2]

1 Computer Science Department, Vrije Universiteit Amsterdam, Amsterdam, The Netherlands

2 Department of Clinical Neuroscience, Karolinska Institutet, Stockholm, Sweden

* Equal contribution

Corresponding authors: a.yaman@vu.nl (AY) & bjorn.lindstrom@ki.se (BL)

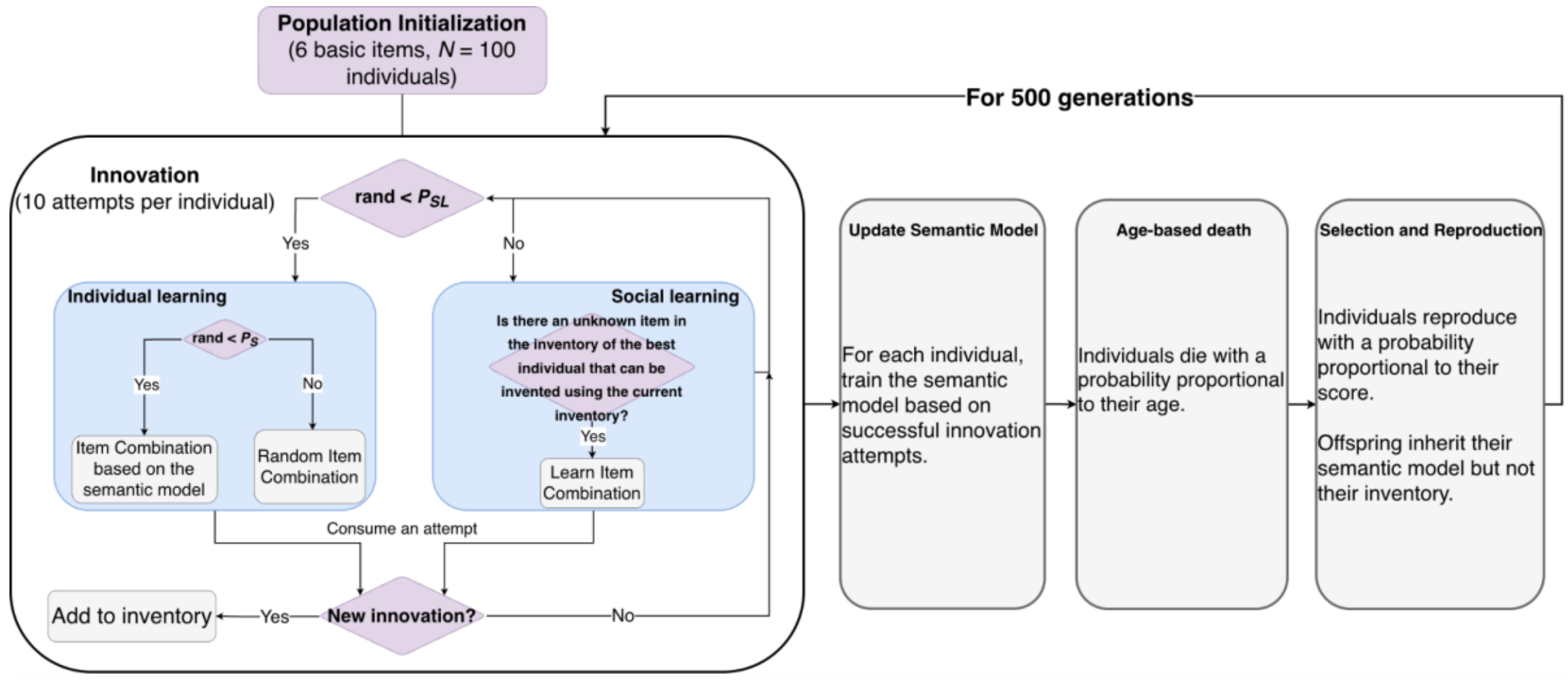

**Figure S1: Overview of the agent-based simulation.** Our model of CCE is based on simulating populations of individuals using a Moran-style selection process in a combinatorial innovation task. See main text for details. $P_{SL}$ = probability of social learning, $P_S$ = probability of using the semantic model, and rand = uniformly distributed random number.

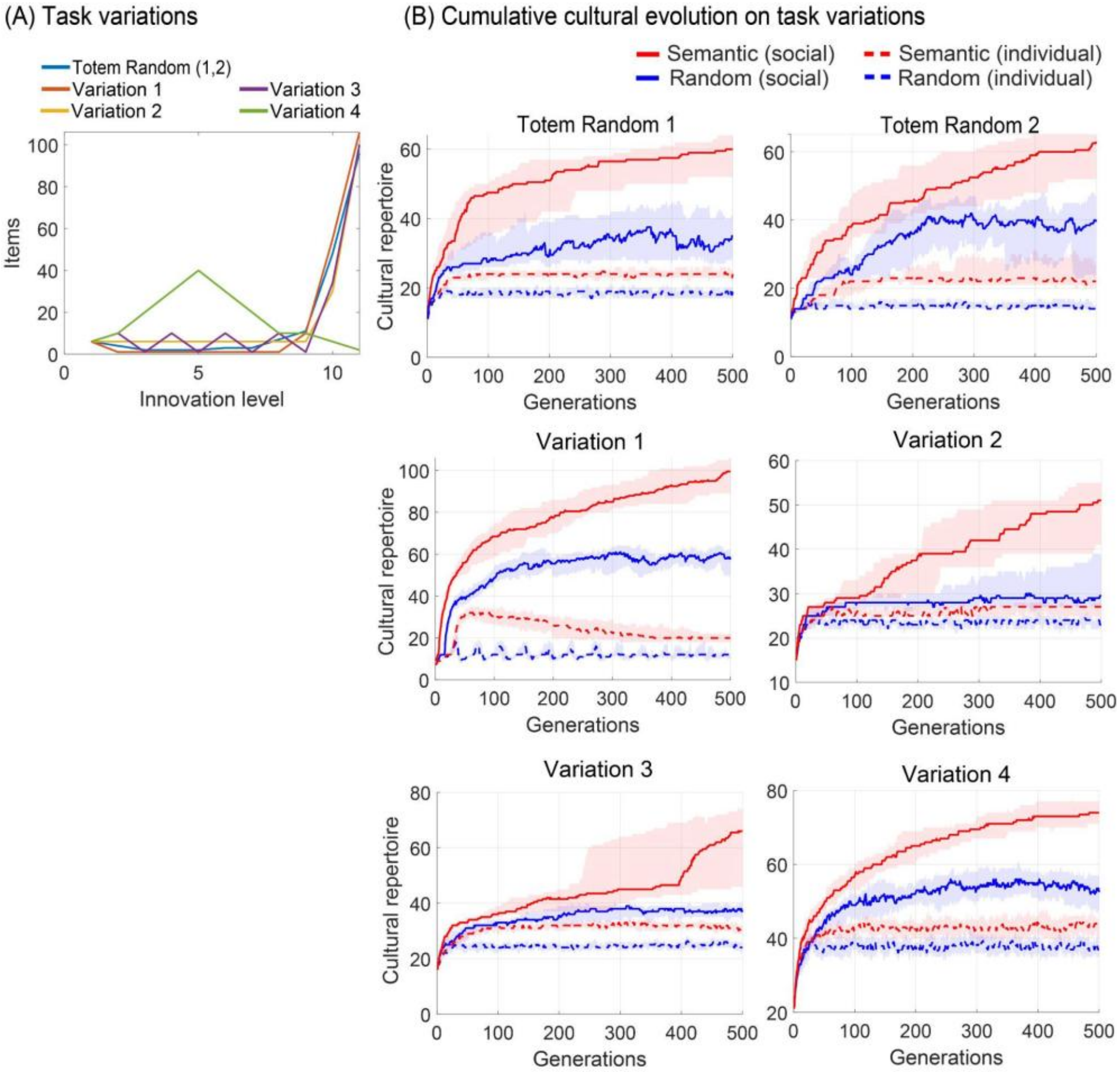

**Figure S2. Sensitivity analysis: Effect of innovation task structure variations on CCE.** (a) Alternative task structures were generated by randomly altering innovation recipes. Each task was hierarchically organized by innovation level, starting with six initial items, and subsequent levels were derived from prior combinations. The baseline *Totem* task included 11 innovation levels (6, 4, 2, 2, 2, 3, 3, 7, 11, 48, 96 items in each level). *Totem Random 1* and *2* retained the same number of items per level but varied the combination recipes, while *Variations 1–4* modified both item counts and recipes. For example, *Variation 1* reduced early-level items and increased later ones, whereas *Variation 2* did the opposite; *Variation 3* introduced highly uneven distributions (e.g., 6, 10, 1, 10, 1, 10, 1, 10, 1, 34, 100), and *Variation 4* increased lower-level items while reducing later ones. (b) Across all variants, populations equipped with both semantic knowledge and social learning consistently achieved the largest cultural repertoires. The *random social* condition outperformed the *semantic individual* condition, highlighting the continued advantage of social learning. Compared to the baseline *Totem* task, *Totem Random 1* and *2* produced smaller repertoires, indicating that altering innovation recipes while preserving item counts increased task difficulty. Similarly, tasks with more items in early innovation levels (e.g., *Variations 2* and *3*) yielded smaller repertoires, reflecting the greater combinatorial search space. *Variation 3* produced the smallest repertoires overall, suggesting it was the most difficult task structure. Lines represent the median of 32 runs, with shaded areas indicating the interquartile range (1st–3rd quartiles).

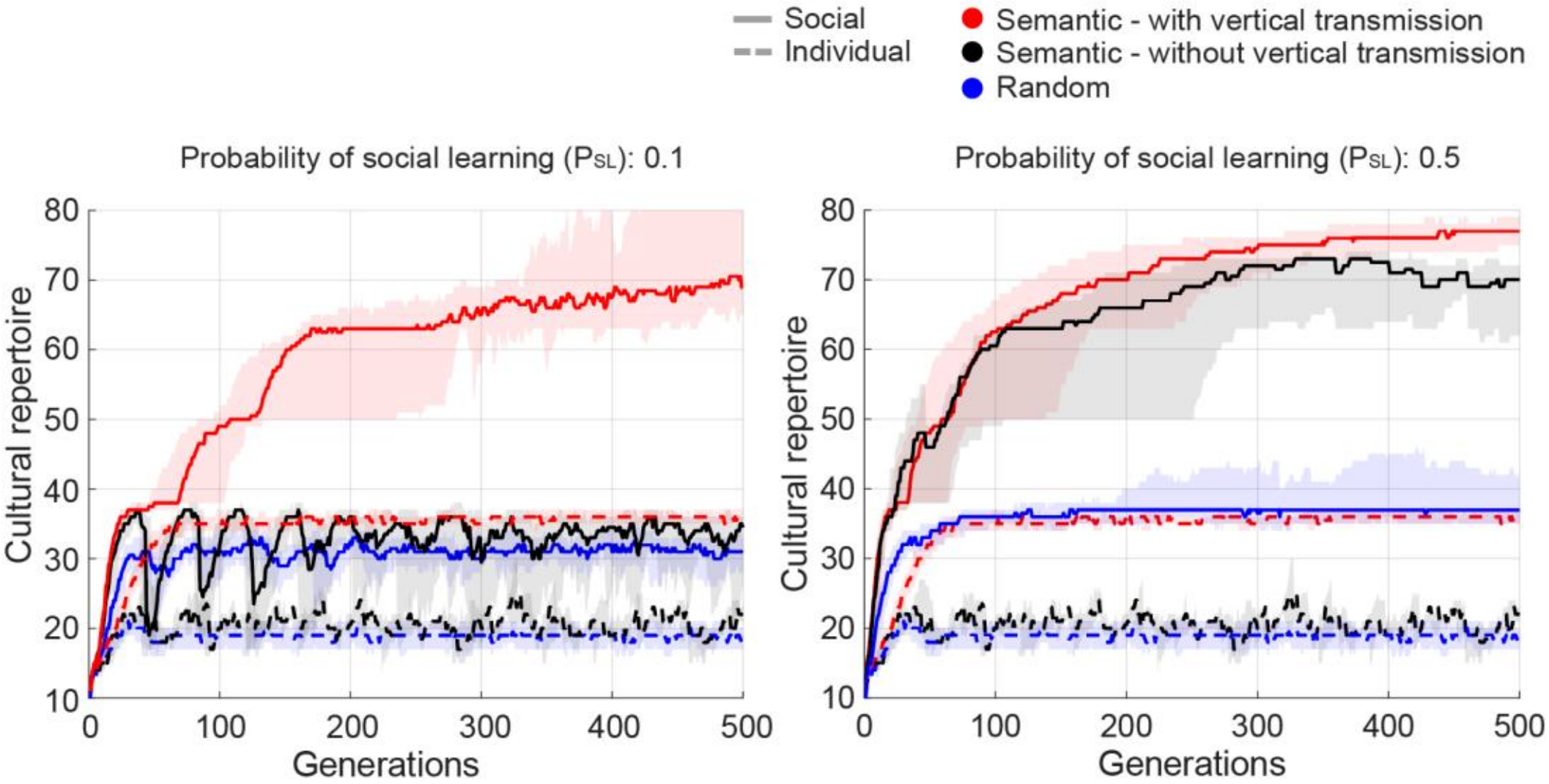

**Figure S3. Sensitivity analysis: Effect of vertical transmission of the semantic model on CCE.** The cultural repertoires of two types of semantic populations—those with and those without vertical transmission—were compared. This comparison was performed in populations of 100 individuals engaging in both individual and social learning. For reference, random populations using individual and social learning were also included. Lines represent the median of 32 runs, with shaded areas indicating the interquartile range (1st–3rd quartiles).

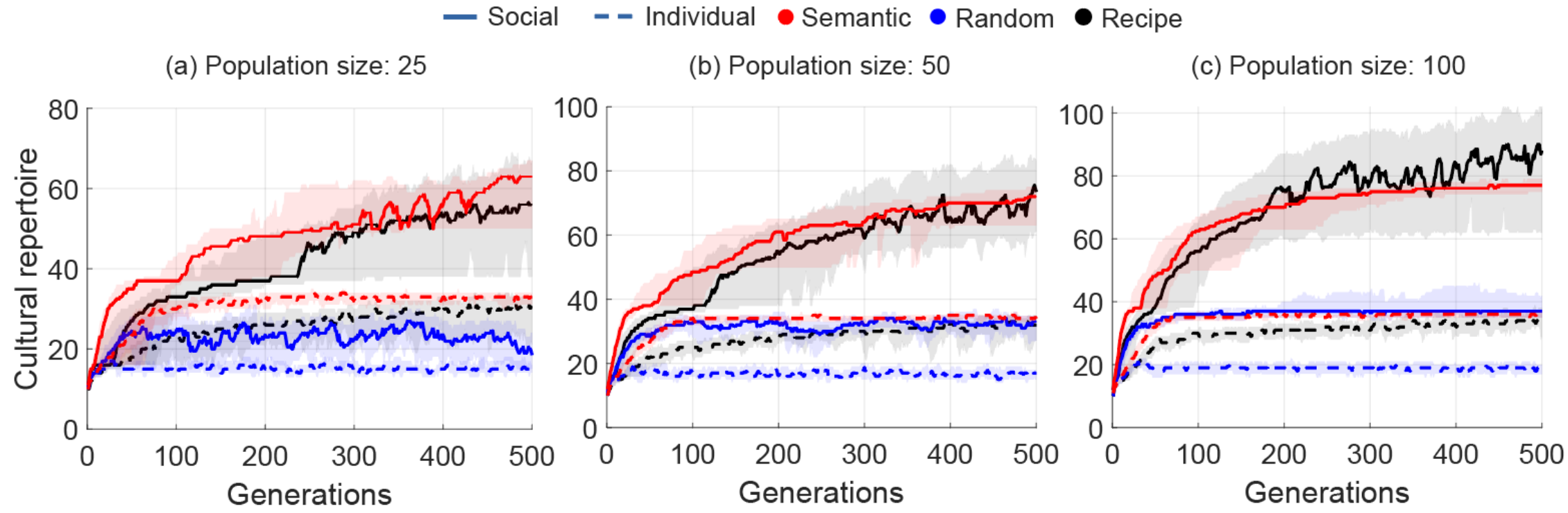

**Figure S4. Sensitivity analysis: Effect of vertical transmission mechanisms on CCE.** The cultural repertoires of populations utilizing two distinct vertical transmission mechanisms—semantic transmission and recipe transmission—were compared. This comparison was performed in three different population sizes (25, 50, and 100) and included both individual and social learning. Additionally, random populations using individual and social learning were also included. Lines represent the median of 32 runs, with shaded areas indicating the interquartile range (1st–3rd quartiles).

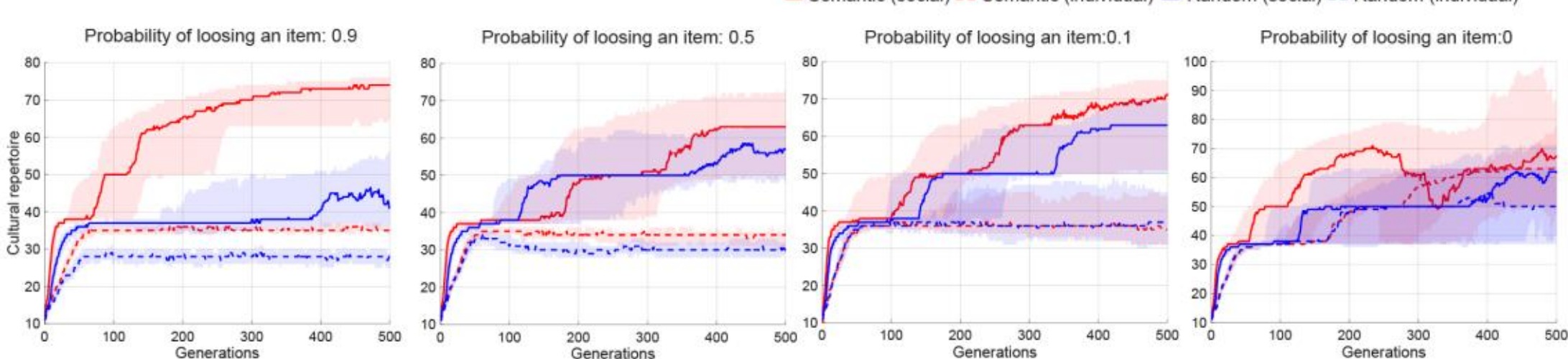

**Figure S5. Sensitivity analysis: Effect of vertical item transmission on CCE.** Cultural repertoire size for populations with and without semantic knowledge under models allowing vertical transmission of discovered items, with a per-item probability of loss during transmission. Item transmission was implemented identically for semantic and non-semantic (random) individuals. While item transmission increased repertoire size for both agent types, populations with semantic knowledge consistently achieved larger repertoires. Item transmission also altered innovation dynamics by enabling individuals to begin life with previously discovered items while reducing opportunities for semantic updating and introducing loss through imperfect transmission and mortality. Non-semantic populations benefited relatively more from item transmission than from knowledge-only transmission. Lines represent the median of 96 simulation runs, with shaded areas indicating the interquartile range (1st–3rd quartiles).

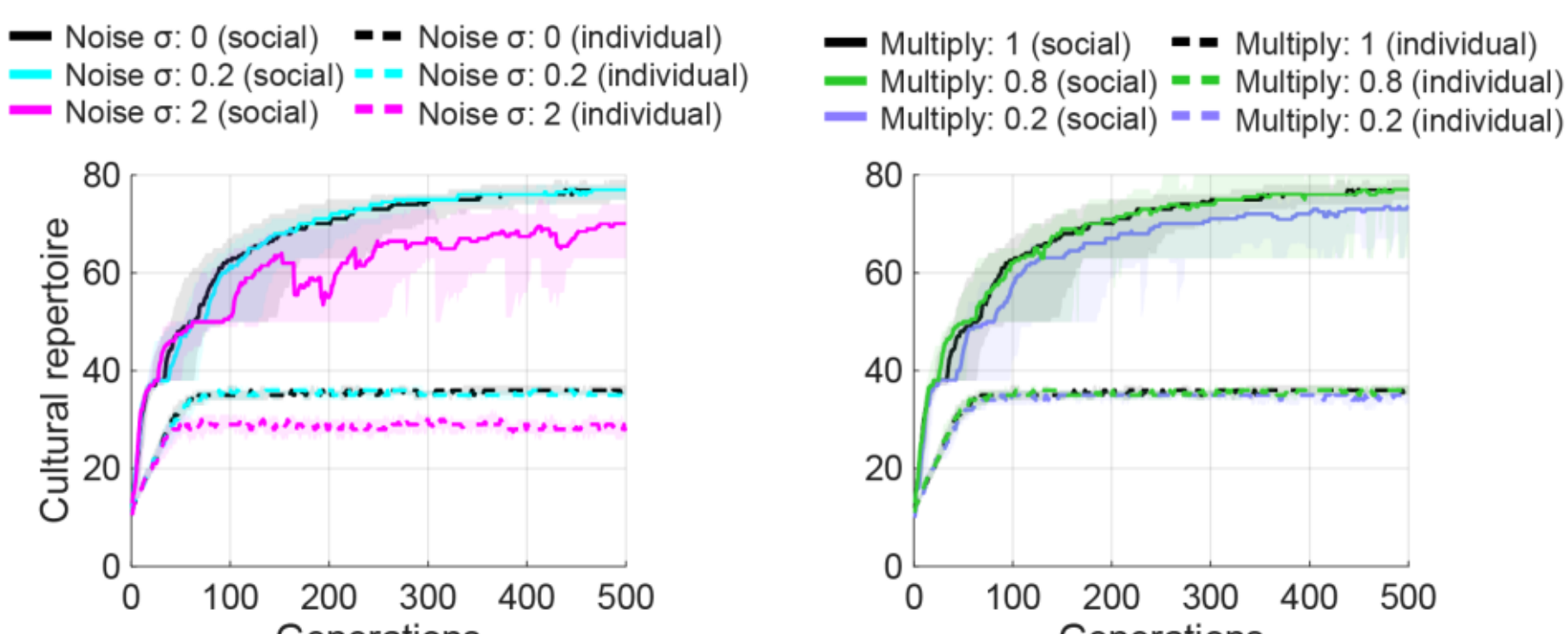

**Figure S6. Sensitivity analysis: The effect of transmission fidelity on CCE.** Cultural repertoire size under two distinct forms of transmission imperfection applied during vertical transmission. (Left) Additive stochastic noise: Gaussian noise (mean = 0; SD = 0.2 or 2) was added to neural network weights and item embeddings (parameters typically ranging from −1 to 1), modeling random transmission error. Cultural repertoires remained robust to moderate levels of noise. (Right) Systematic degradation: Transmission fidelity was manipulated by scalar multiplication of embeddings and neural network parameters, modeling uniform signal attenuation rather than random error. Lines represent the median of 32 runs, with shaded areas indicating the interquartile range (1st–3rd quartiles).

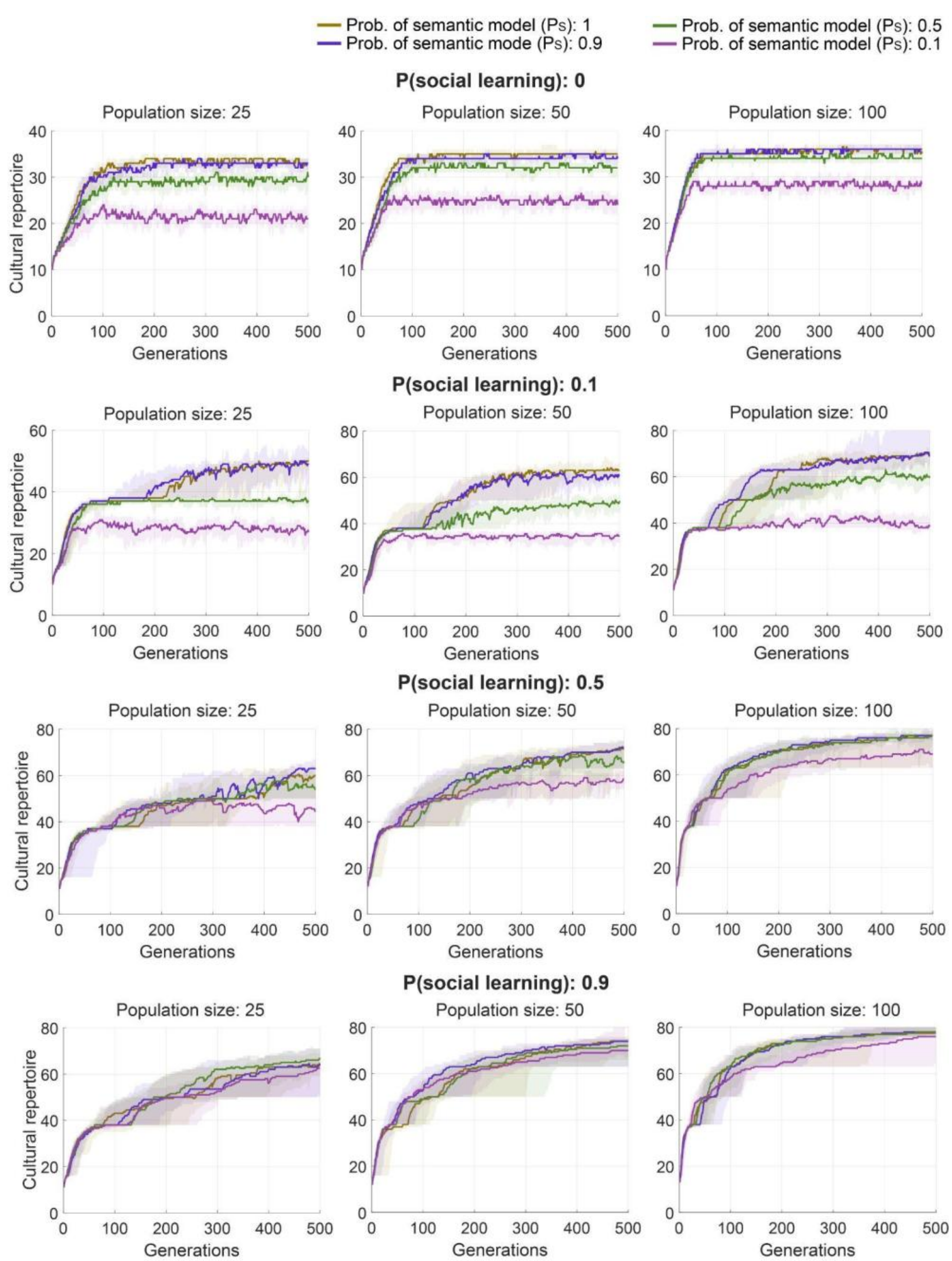

**Figure S7. Sensitivity analysis: Effects of population size and semantic knowledge reliance on CCE.** Cultural repertoire size is shown as a function of population size and the probability of using the semantic model across different levels of social learning probability. Semantic knowledge substantially enhances CCE, even when social learning is infrequent. Social learning provides successful examples that improve individuals' semantic models, amplifying innovation across population sizes and levels of semantic knowledge. Lines indicate the median of 32 runs; shaded areas denote the interquartile range (1st–3rd quartiles).

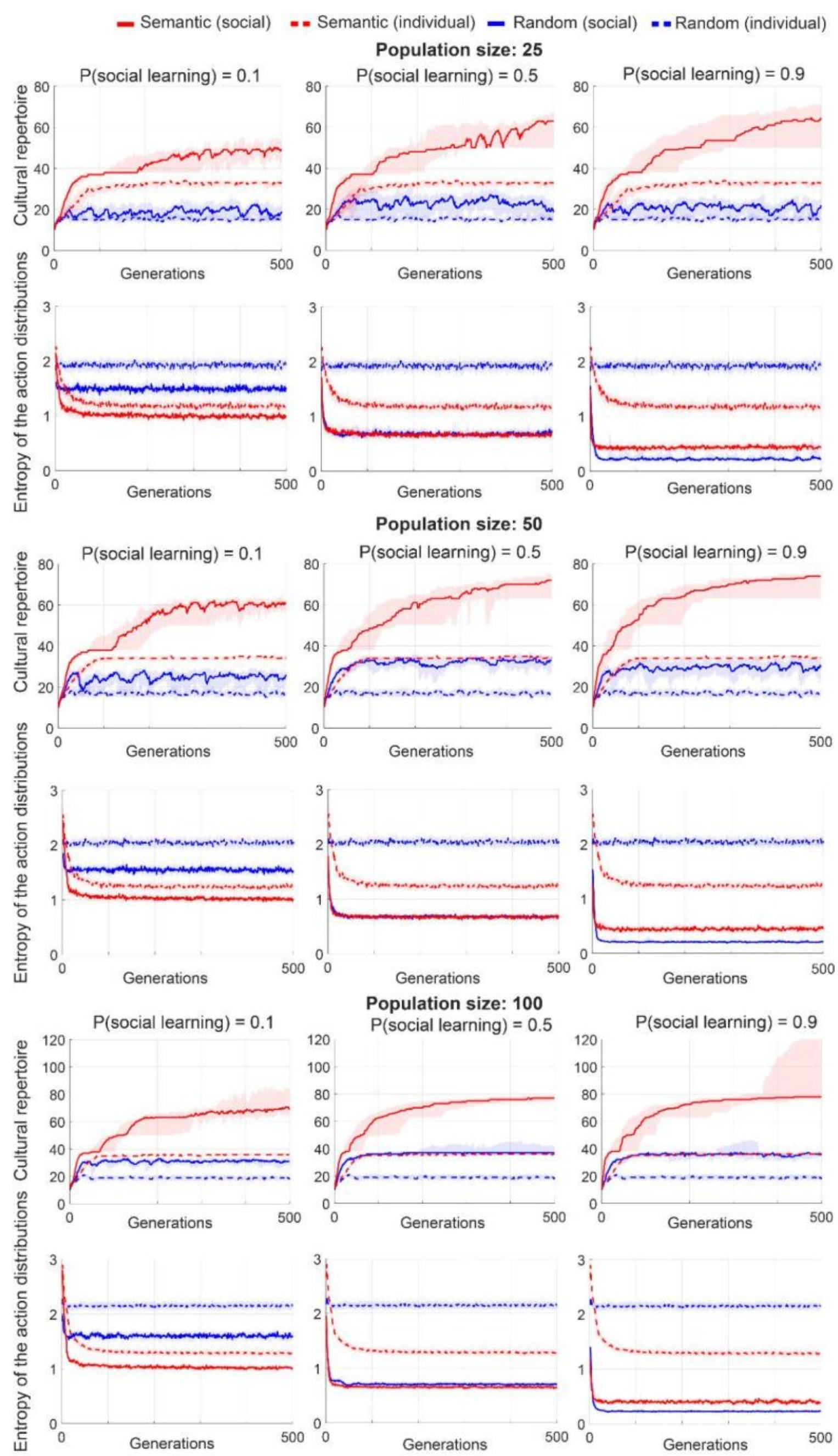

**Figure S8. Sensitivity analysis: Effects of population size and social learning capacities on CCE.** Cultural repertoire size and action distribution entropy are shown as functions of population size and the probability of social learning ($P_{SL}$). Lines indicate the median of 32 runs; shaded areas denote the interquartile range (1st–3rd quartiles).

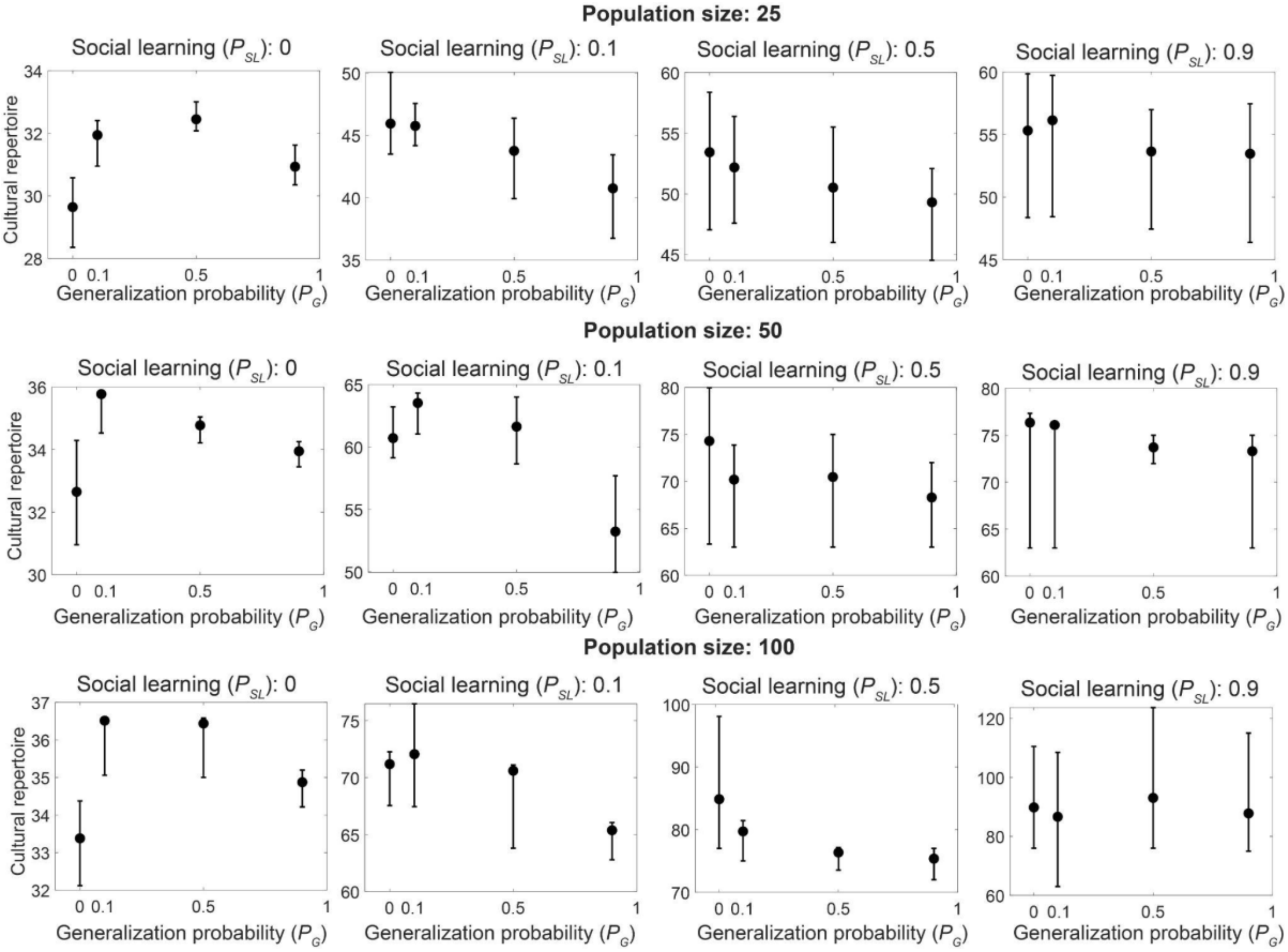

**Figure S9. Sensitivity analysis: Effects of recipe generalization probability under varying population sizes and social learning rates.** Cultural repertoire size as a function of generalization probability, population size, and social learning probability. Without social learning ($P_{SL}$ = 0), moderate generalization consistently boosts innovation, whereas high generalization reduces exploration. With social learning, effects are less consistent: at $P_{SL}$ = 0.1, generalization increases innovation for populations of 50 and 100, but in other cases yields similar or fewer innovations compared to individuals without generalization. These results suggest that in larger populations, collective success benefits from random exploration combined with social learning rather than overgeneralizing from limited experience.

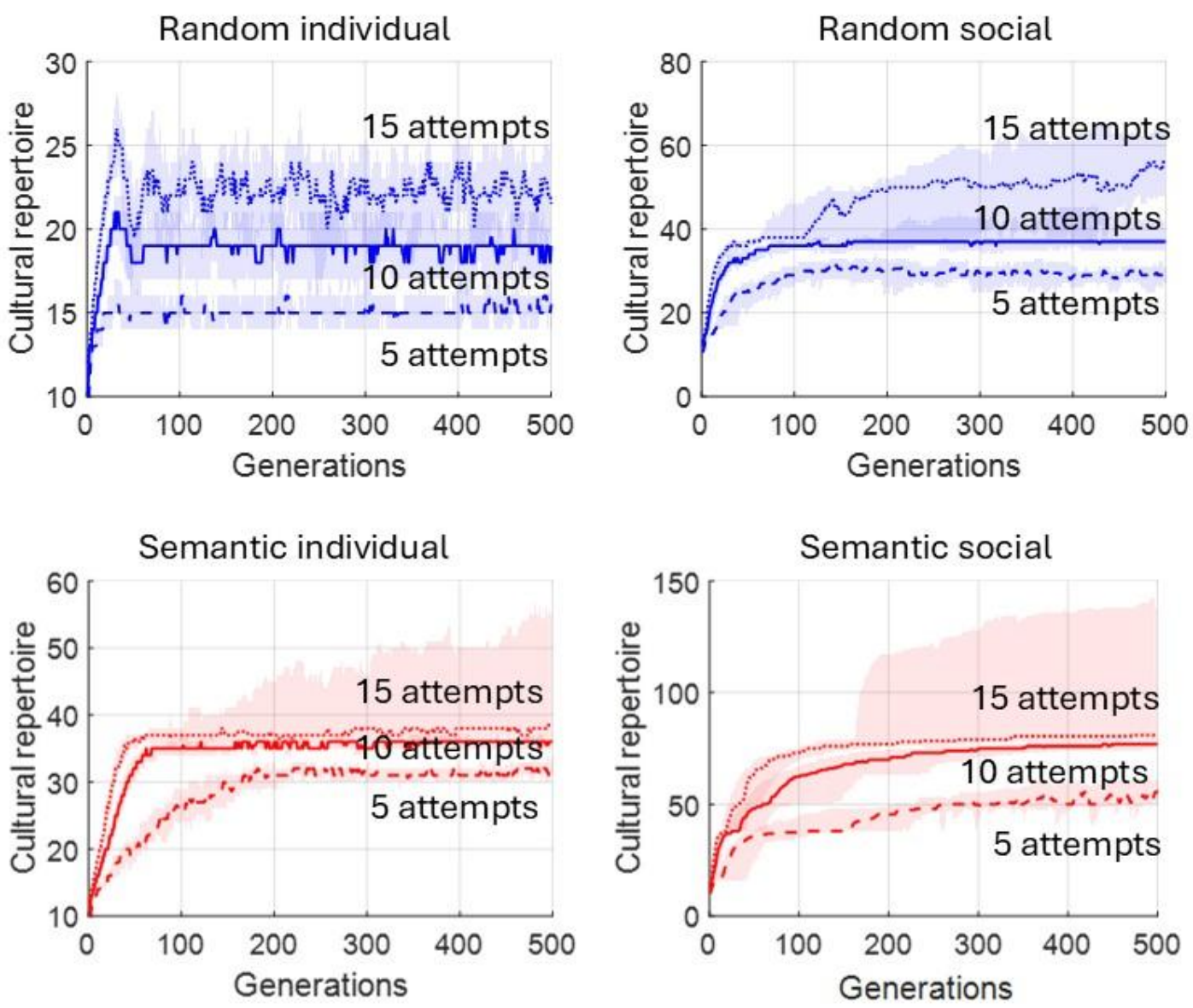

**Figure S10. Sensitivity analysis: Effect of attempt number on CCE.** Cultural repertoire size is shown as a function of the number of attempts individuals can make before reproduction. Allowing more attempts consistently enhanced CCE by increasing exploration opportunities in the non-semantic conditions and providing richer input for semantic updating in the semantic conditions. Lines indicate the median of 32 runs; shaded areas denote the interquartile range (1st–3rd quartiles).

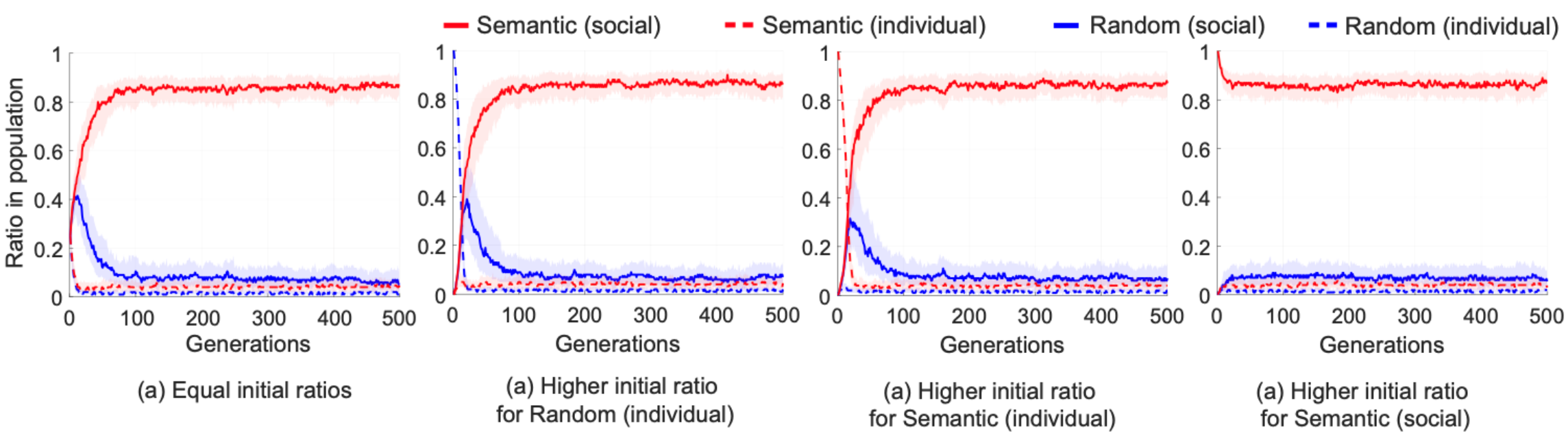

**Figure S11. Sensitivity analysis: Initial strategy proportions.** The figure shows the ratio of different strategies in the population over time, across varying initial proportions of individuals with and without semantic knowledge and social learning. Regardless of starting composition, individuals possessing both semantic knowledge and social learning consistently came to dominate the population. Lines indicate the median of 32 runs; shaded areas denote the interquartile range (1st–3rd quartiles).

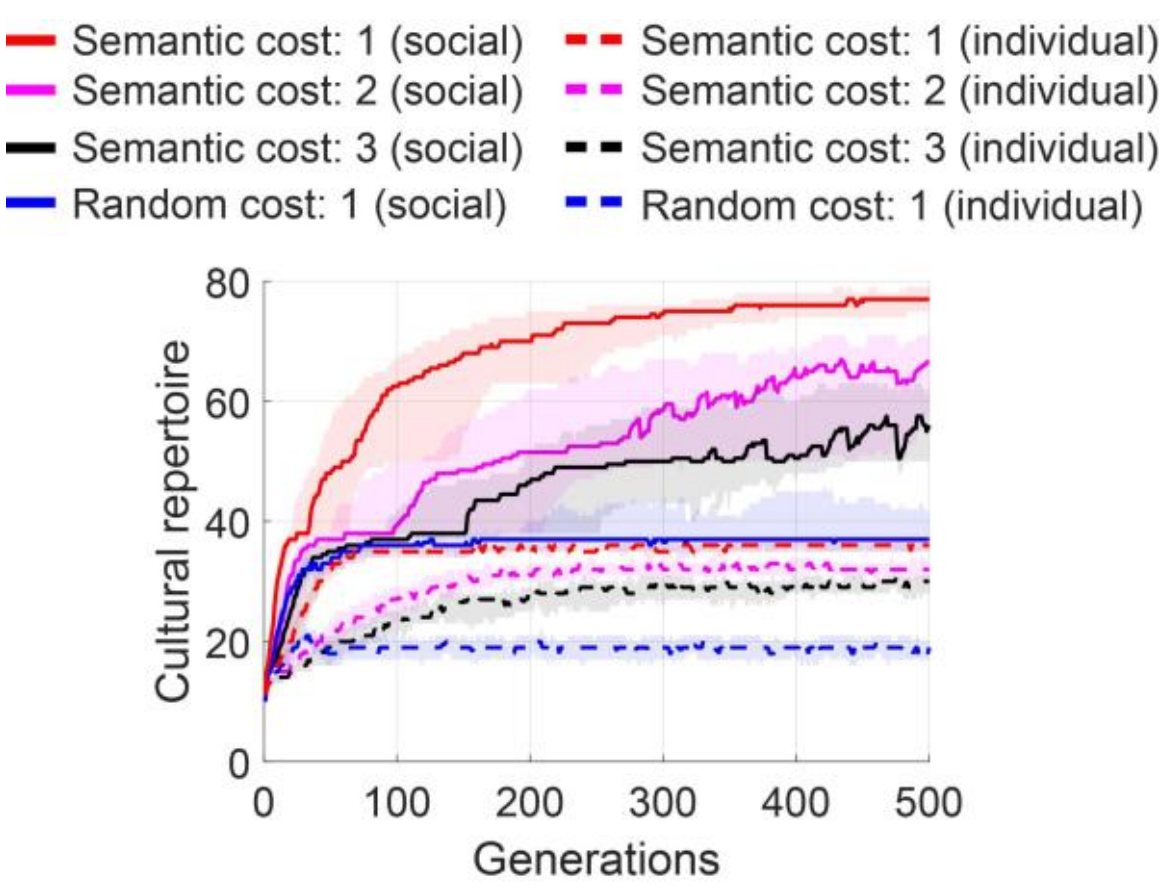

**Figure S12. Sensitivity analysis: Effects of the cost of the semantic model on CCE.** Cultural repertoire size is shown as a function of the cost associated with using the semantic model. The cost is expressed as the number of attempts it consumes per generation (out of a maximum of 10). In baseline simulations, the cost was 1 (no additional cost). Results show that even when using the semantic model is three times as costly as no model, it remains beneficial for CCE. Lines indicate the median of 32 runs; shaded areas denote the interquartile range (1st–3rd quartile).

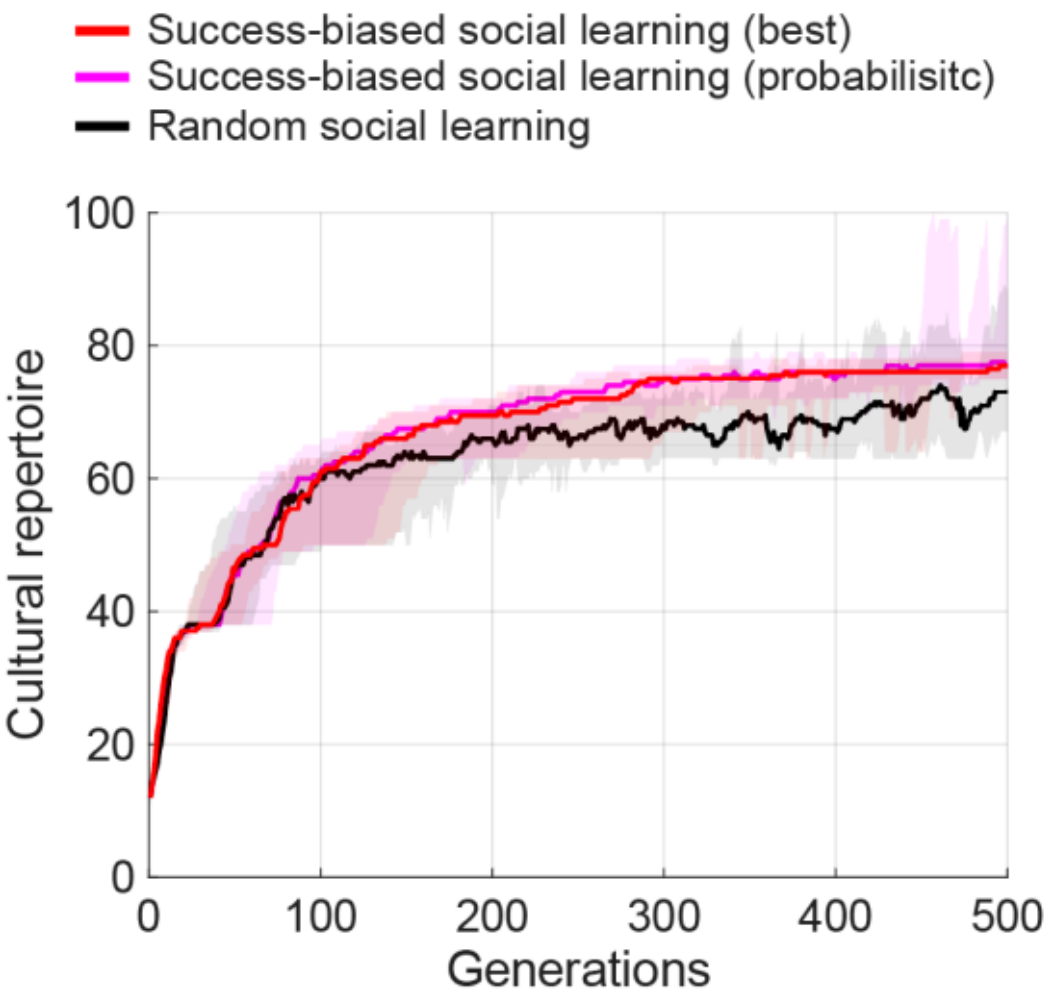

**Figure S13. Sensitivity analysis: Effects of social learning strategies on CCE.** Cultural repertoire size for populations using three social learning strategies: deterministic success-biased (copying the highest-scoring individual), probabilistic success-biased (copying individuals with probability proportional to their score), and random (copying a randomly selected individual). In all simulations, the probability of social learning was fixed at $P_{SL} = 0.5$. Lines show the median across simulation runs; shaded areas indicate the 1st and 3rd quartiles.

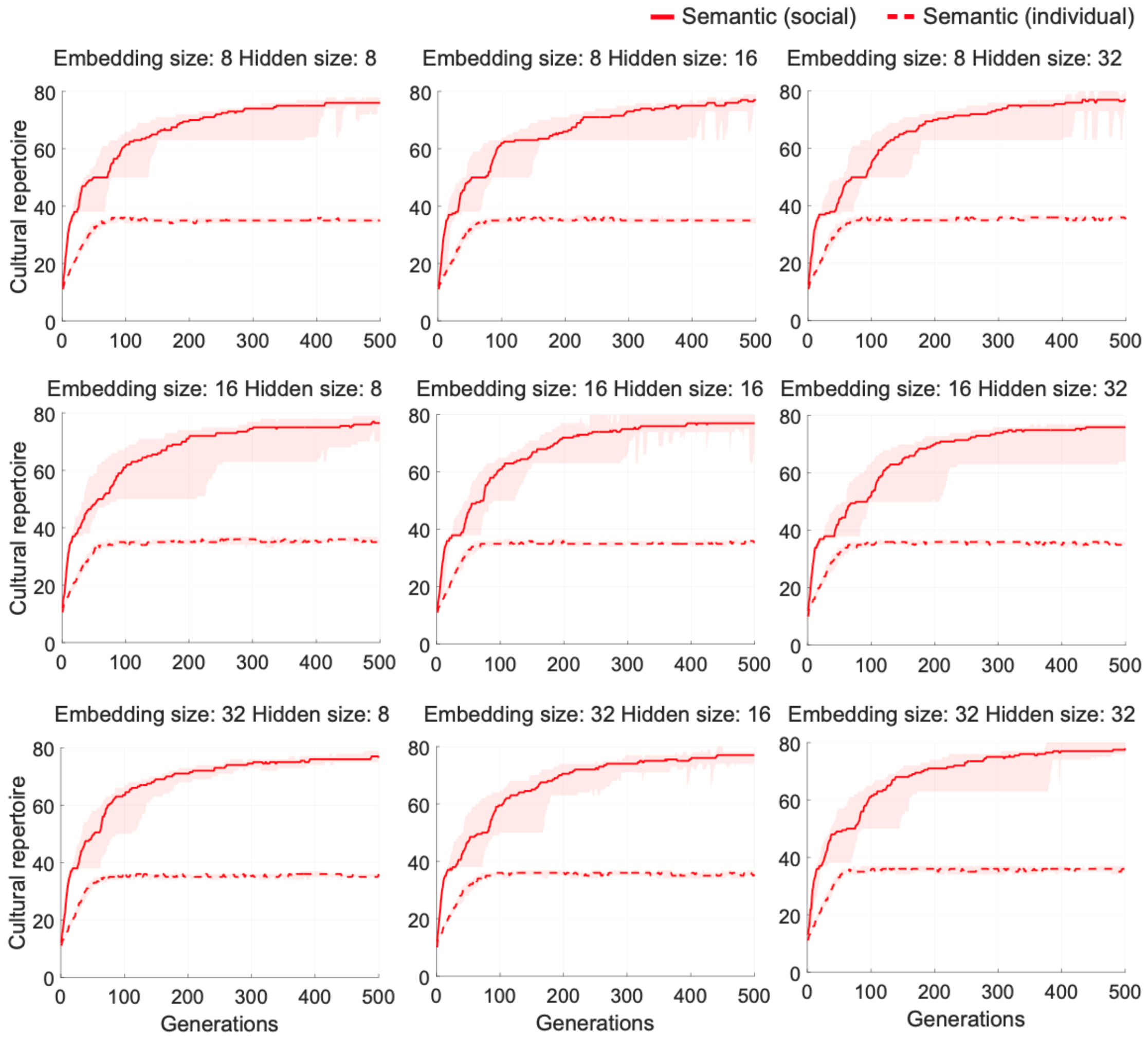

**Figure S14. Sensitivity analysis: Effect of neural network hyperparameters on CCE.** Cultural repertoire size as a function of embedding size and number of hidden-layer neurons. Across all configurations, conditions with semantic knowledge consistently produced larger cultural repertoires, confirming that our findings do not depend on specific neural network hyperparameter choices. Lines indicate the median of 32 simulation runs, and shaded areas show the 1st and 3rd quartiles.

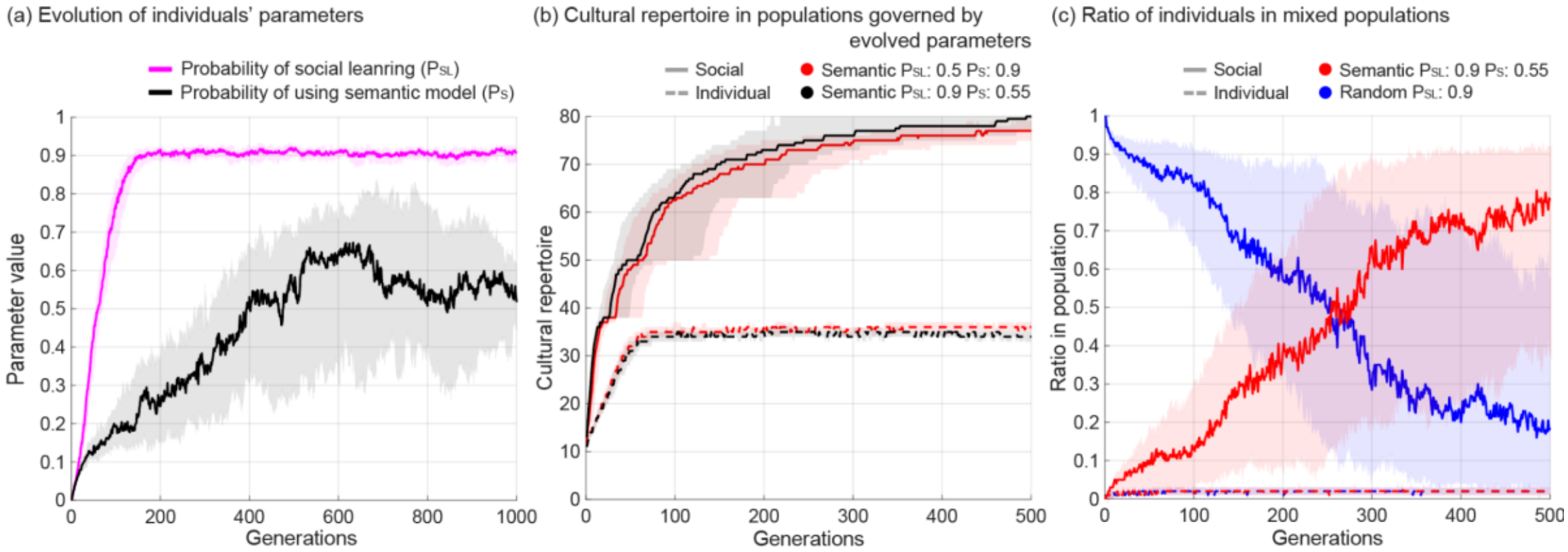

**Figure S15. Evolution of social learning and semantic model use. (a)** Evolution of the model parameters governing the probability of social learning ($P_{SL}$) and the probability of using the semantic model during individual exploration ($P_S$). Starting from initial values of zero, $P_{SL}$ converges to approximately 0.9 and $P_S$ to approximately 0.55. **(b)** Individuals with evolved parameter values produce comparable cultural repertoires to the main results (Figure 2a). **(c)** When both capacities are allowed to evolve within mixed populations, the individuals with the optimized parameters ultimately become dominant, which is consistent with Figure 2b. Lines represent the median of 32 runs, with shaded areas indicating the interquartile range (1st–3rd quartiles).

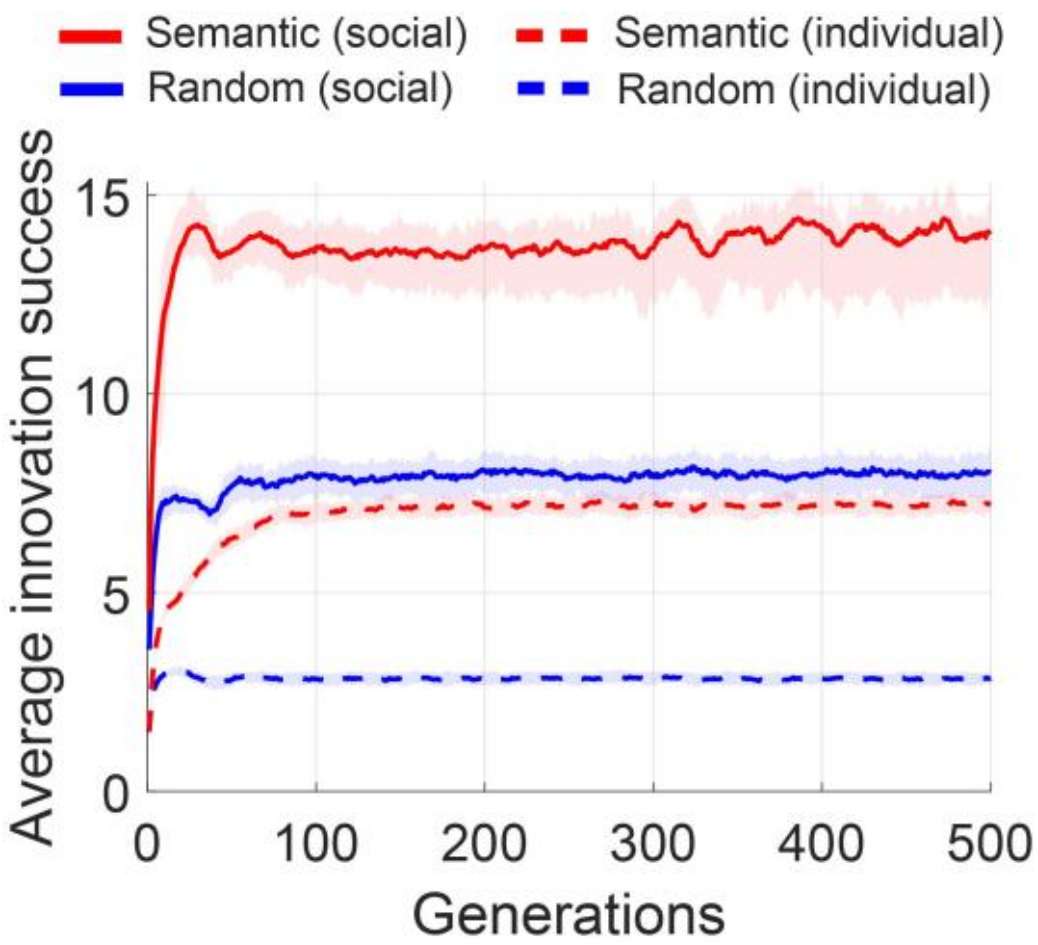

**Figure S16. Average innovation success**. Average innovation success across generations in populations of 100 individuals, comparing conditions with and without semantic and social learning capacities. Unlike the population cultural repertoire shown in Figure 2a, there is no clear synergy between semantic knowledge and social learning. Lines indicate the median across 96 simulation runs, and shaded areas represent the 1st and 3rd quartiles.

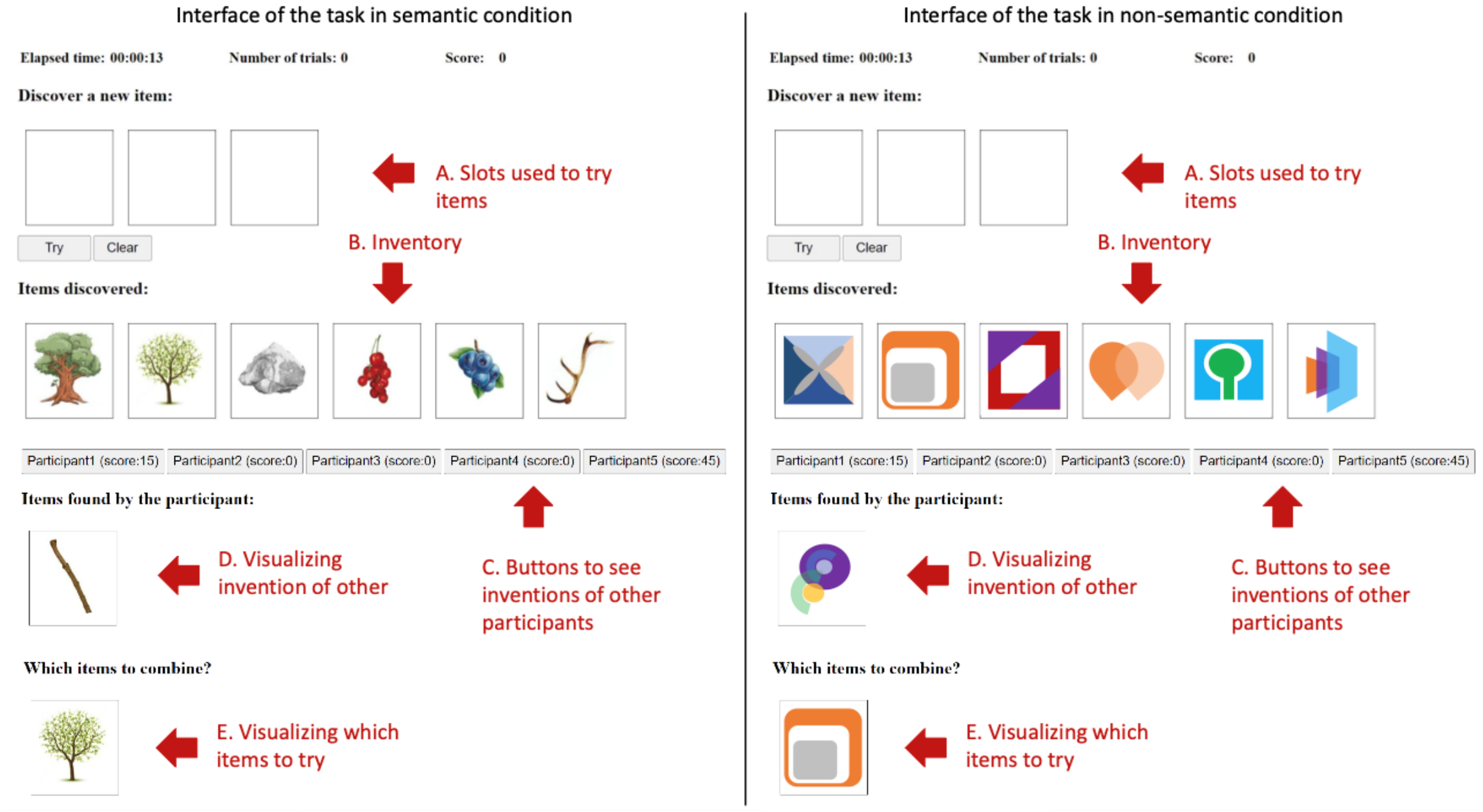

**Figure S17. The interface of the innovation task used in the experiment.** The left panel shows the version used in the semantic condition, while the right panel shows the version used in the non-semantic condition. Although redrawn for copyright purposes, the abstract symbols closely mirror those used in the actual experiment.

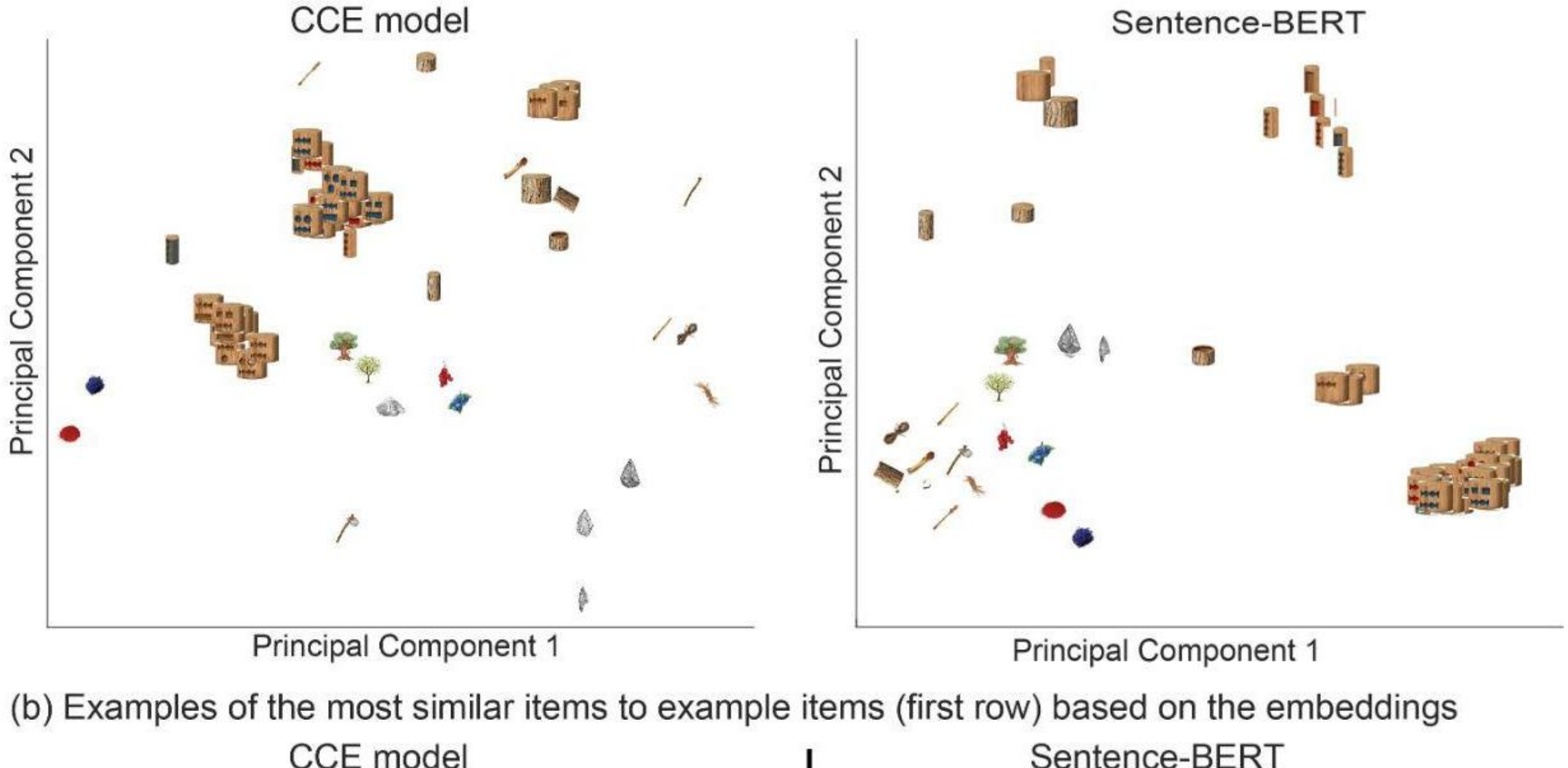

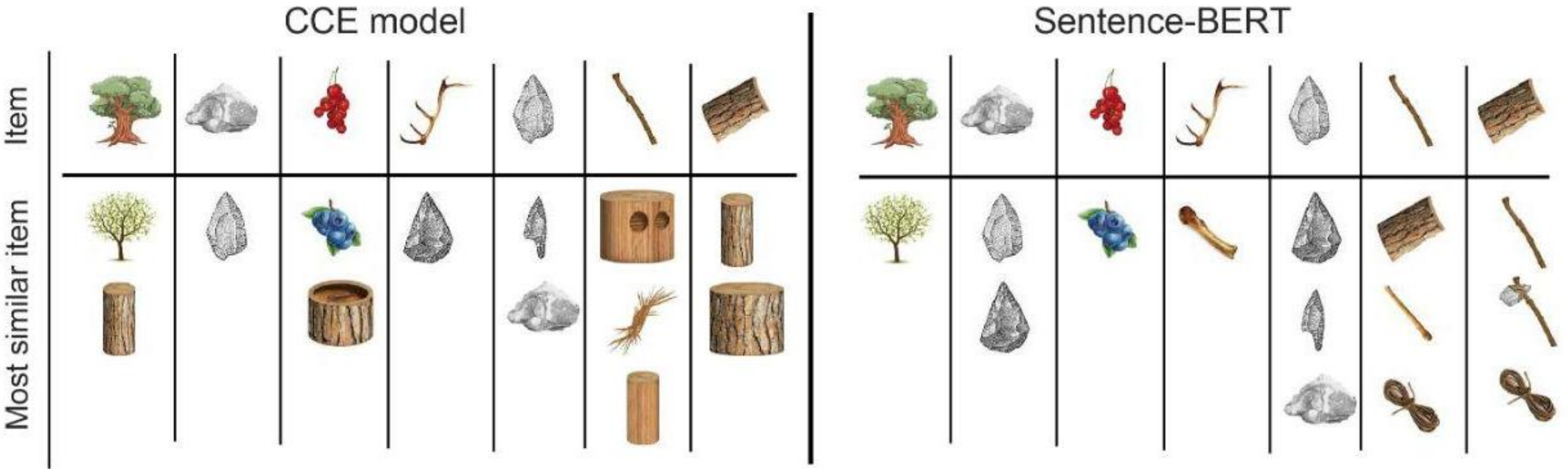

**Figure S18. Comparisons between semantic representations learned from the CCE model and those derived from the Sentence-BERT model. (a)** Principal Component Analysis (PCA) was applied to the similarity matrices generated from both the model-learned vector representations and the sentence embeddings. The first two principal components for each item are visualized. In both cases, semantically similar items formed distinct clusters. Qualitatively, the clustering pattern observed in the representations learned through our evolutionary simulation closely resembles those produced by Sentence-BERT, suggesting comparable underlying semantic structure. **(b)** Our CCE model and the Sentence-BERT model represent the semantic distance between items in a similar manner.

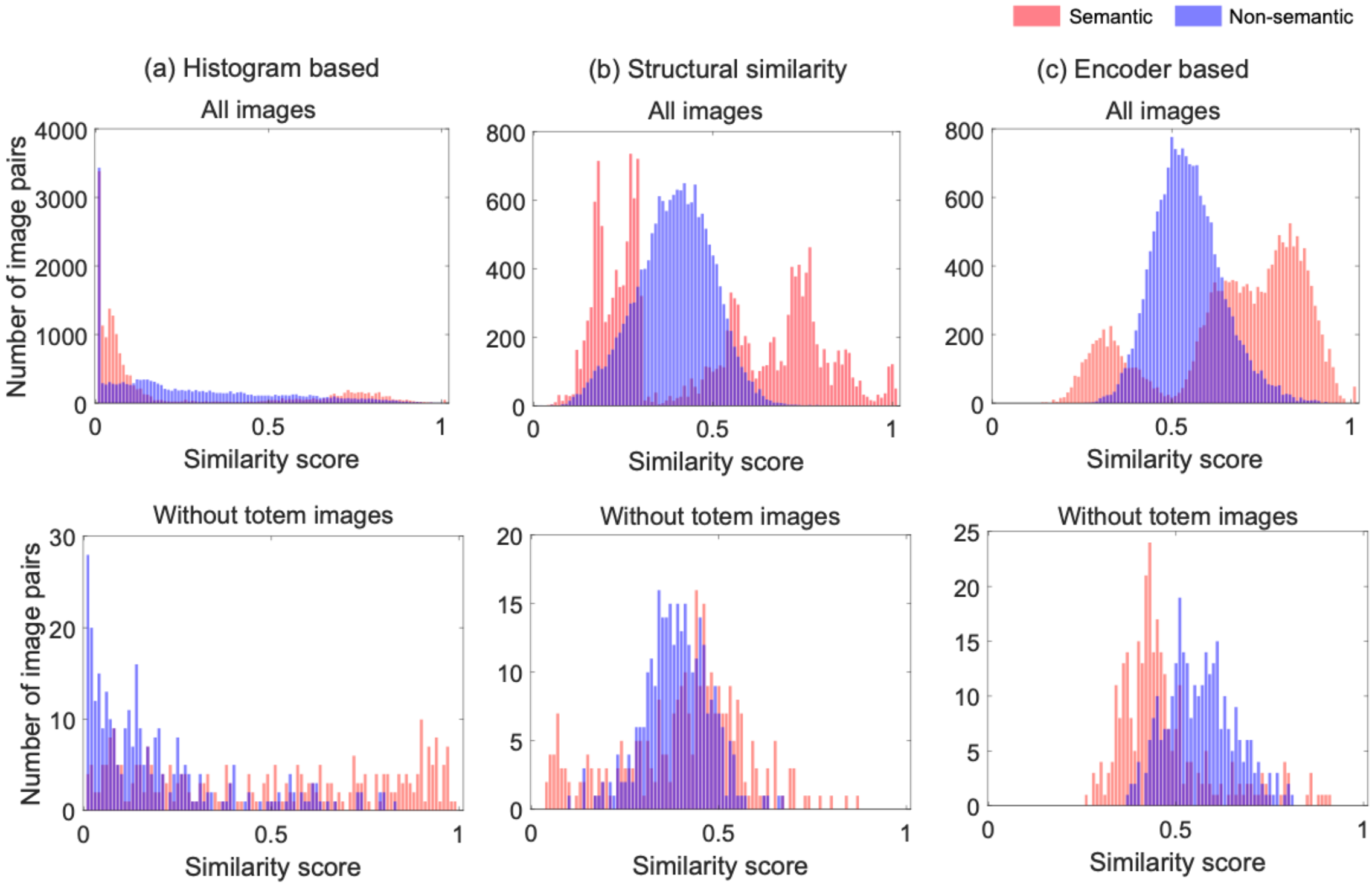

**Figure S19. Three image similarity metrics.** Although the images used in the non-semantic condition are abstract symbols, they show lower similarity compared to the semantically meaningful images. We measured these similarities using three distinct metrics: **(a)** histogram-based, **(b)** structural similarity, and **(c)** deep learning-based encoder similarity. Since many items in the semantic condition were visually similar totems, we computed image similarities both with and without these totems. In both cases, the non-semantic images showed lower similarity across the metrics, indicating that they were sufficiently distinct for human participants to differentiate between them.

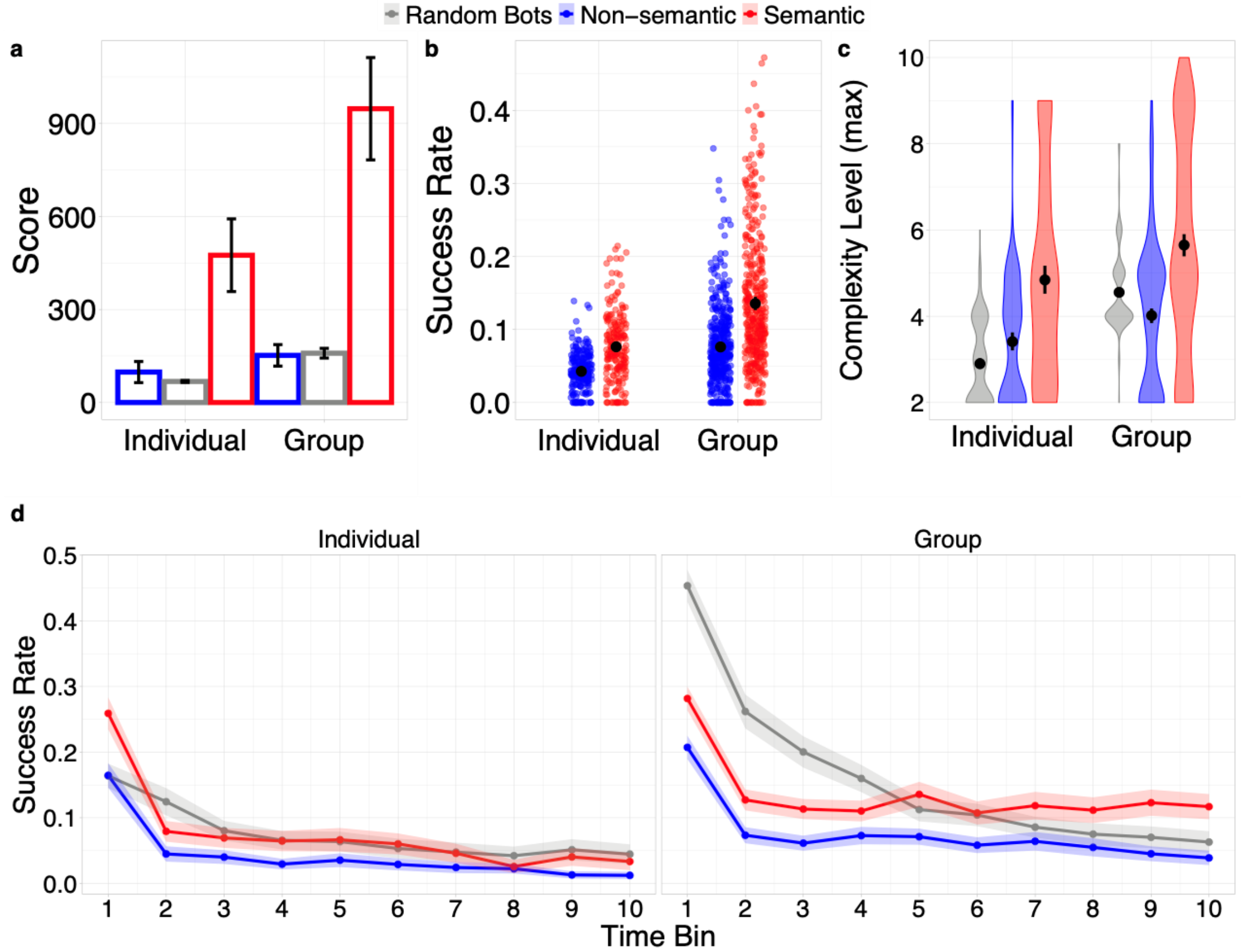

**Figure S20. Additional task performance results. (a)** Average scores achieved across all conditions, with 95% confidence intervals. The results are consistent with those of the original study using the semantic condition, comparing random agents and human participants in individual and group conditions (2). **(b)** These findings were replicated when using the success rate, defined as the number of innovations divided by the number of attempts. **(c)** The maximum level of complexity achieved by each participant. **(d)** The success rate was calculated as the proportion of successful innovations within each of 10 equally-sized time bins per participant. To account for the increasing difficulty of discovering new items as the inventory expands, an adjusted success rate was computed by multiplying the raw success rate by a difficulty factor. The difficulty factor is the ratio of the current action space to the baseline action space, with 6 starting items. Thus, maintaining the same raw success rate with a larger inventory yields a higher adjusted success rate, reflecting better performance relative to the expanded search space. Shaded regions indicate 95% confidence intervals. Solid and dashed lines represent individual and group conditions, respectively.

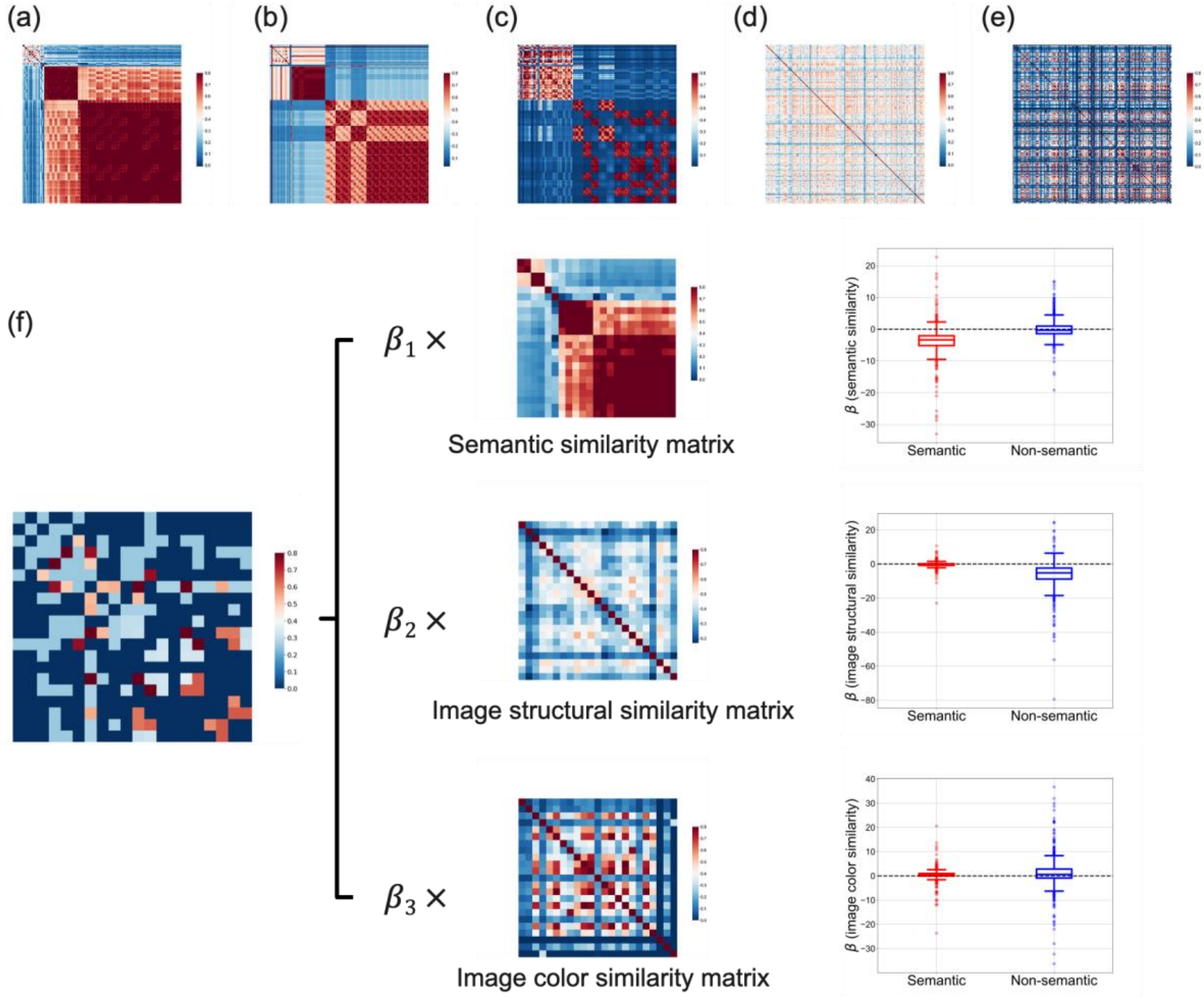

**Figure S21. Behavioral Representational Similarity Analysis: procedure and results. (a)** Semantic similarity matrix showing pairwise similarities between all items in the semantic condition. **(b)** Structural image similarity matrix for items in the semantic condition, based on structural visual features. **(c)** Color distribution similarity matrix for items in the semantic condition. **(d)** Structural image similarity matrix for items in the non-semantic condition. **(e)** Color distribution similarity matrix for items in the non-semantic condition. **(f)** Participants' actions were influenced by semantic dissimilarity in the semantic condition, whereas in the non-semantic condition, behavior was more strongly driven by visual structural dissimilarity. This analysis also controls for the contribution of color dissimilarity.

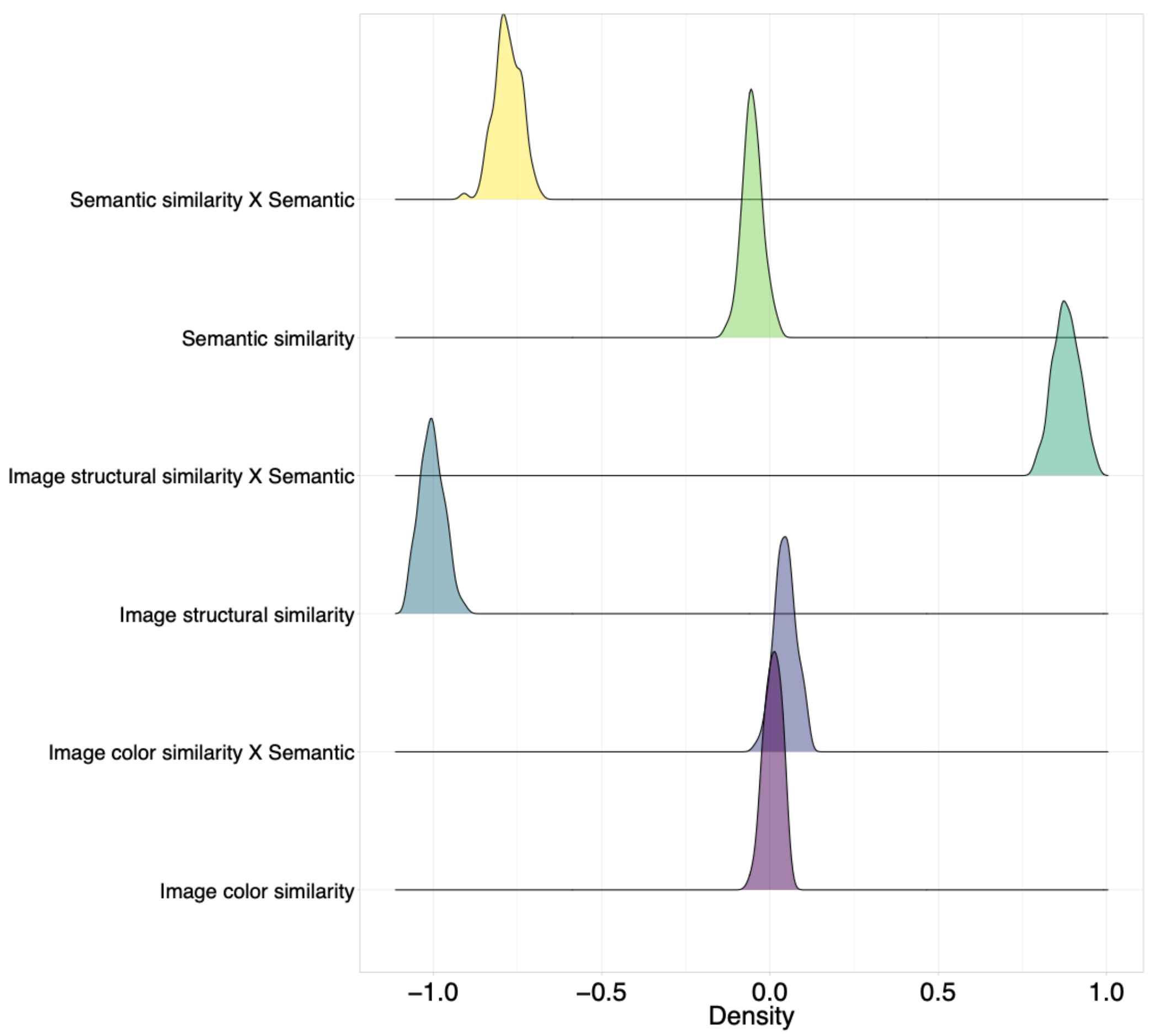

**Figure S22. Distribution of model estimates of different behavioral strategies.** To account for the stochasticity of alternative actions sampled from the entire action space, we ran the GLMM 100 times with different alternative actions. The figure depicts the distribution of beta estimates for each feature. The coefficients for each feature represent the non-semantic (baseline) condition, while the interaction coefficients represent the difference between the semantic and non-semantic conditions.

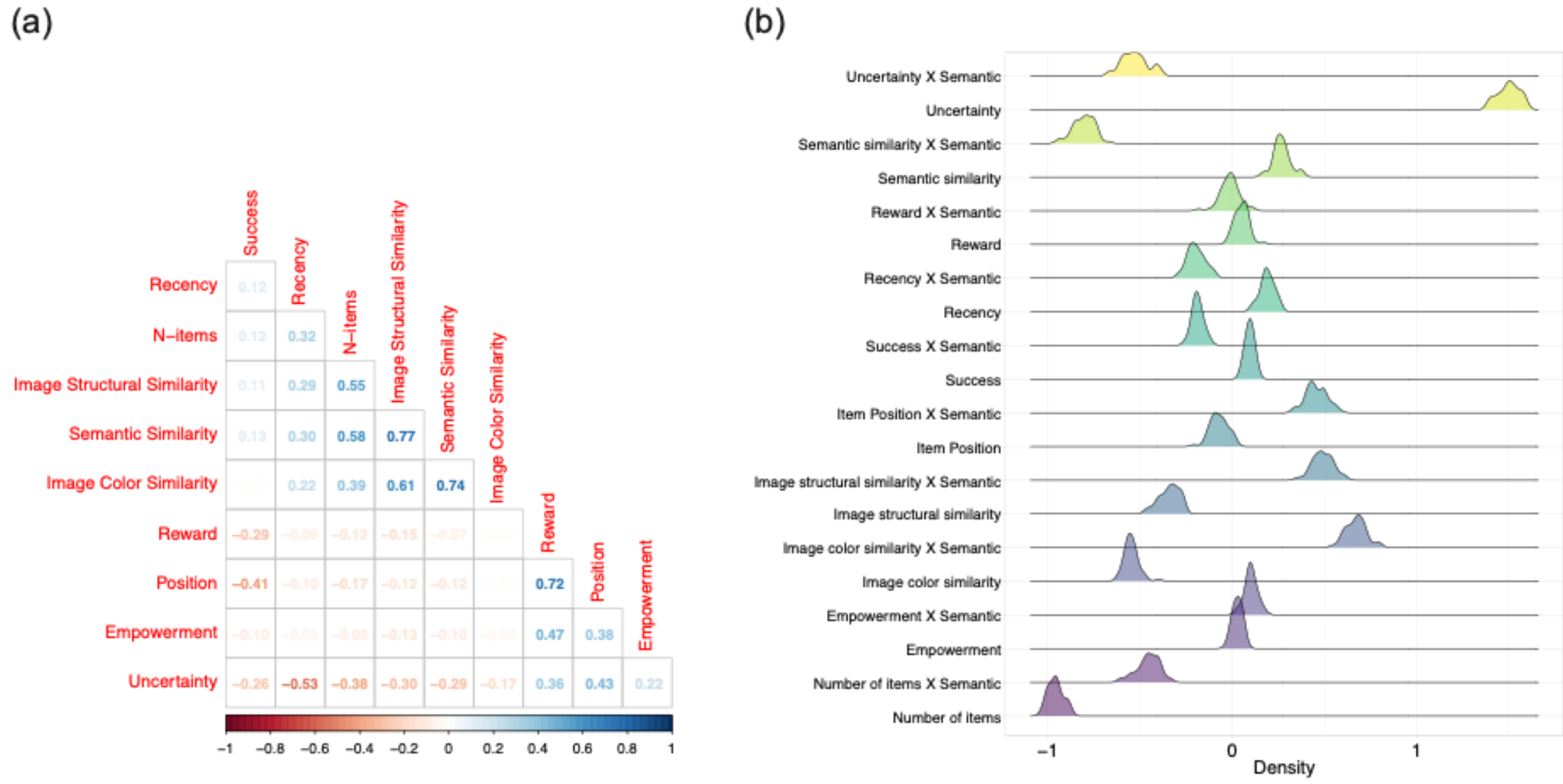

**Figure S23. Distribution of model estimates of all behavioral strategies.** To account for the stochasticity of alternative actions sampled from the entire action space, we ran the model 50 times with different randomly sampled alternative actions. The figure depicts the **(a)** average correlation between the features, and **(b)** the distribution of beta estimates for each feature. The coefficients for each feature represent the non-semantic (baseline) condition, while the interaction coefficients represent the difference between the semantic and non-semantic conditions.

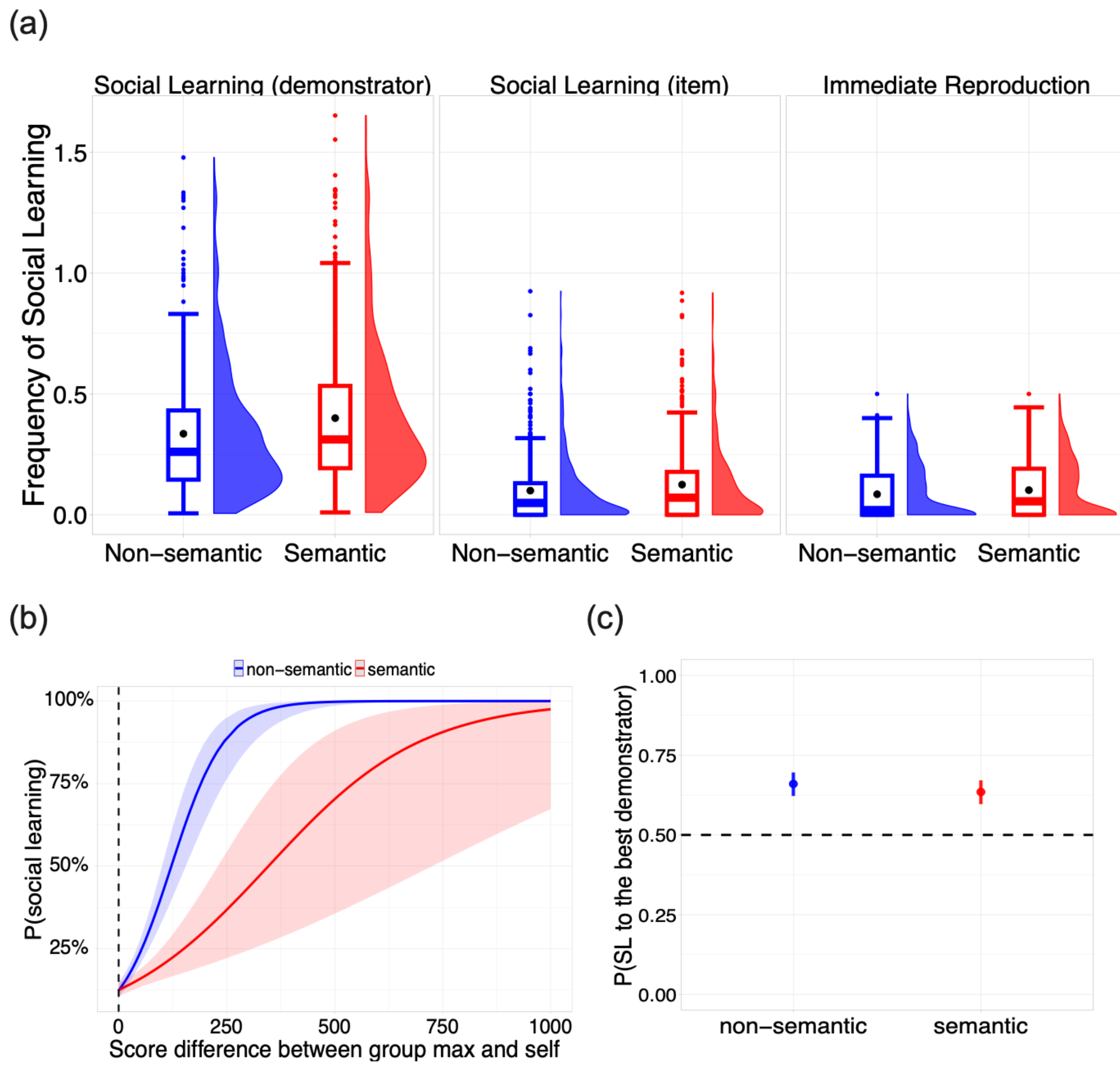

**Figure S24. Experimental results: Additional analysis of social learning.** (a) Three indices of social learning were examined: the frequency of inspecting other group members' inventories, the frequency of inspecting specific items, and the frequency of immediately reproducing the inspected item in the following attempt. No significant differences in overall social learning tendencies were found between the two conditions. (b) Participants were more likely to engage in social learning when there was a larger gap between their own score and the highest score in the group; this effect was particularly pronounced in the non-semantic condition. (c) Participants strategically learned from the group member with the highest score.

| Parameter | Meaning | Value |
| --- | --- | --- |
| | Number of neurons in hidden layers | {8, **16**, 32} |
| | Size of the item embeddings | {8, **16**, 32} |
| $P_D$ | Probability of death | see Alg. 1 |
| $P_S$ | Probability of using the semantic model | {0, 0.1, 0.5, **0.9**} |
| $P_{SL}$ | Probability of using social learning | {0, 0.1, **0.5**, 0.9} |
| $P_G$ | Probability of using generalization | {0, **0.1**, 0.5, 0.9} |

**Table 1. Individual model parameters.** The table displays parameter names, their meaning, and the range of values evaluated in the sensitivity analysis. The parameter values used for the main text results are bolded.

| Parameter | Meaning | Value |
| --- | --- | --- |
| N | Population size | {25, 50, **100**} |
| $n_{attempt}$ | The number of attempts per generation | {5, **10**, 15} |

**Table 2. Global model parameters.** The table displays parameter names, their meaning, and the range of values evaluated in the sensitivity analysis. The parameter values used for the main text results are indicated in bold type.

**function CCE**
$P_{SL}$: social learning probability
$n_{attempt}$ **:** number of innovation attempts each individual makes in each generation
$P_D$: Death probability based on the equation $P_D = ae^{bx}$ where x is the age of an individual and a: 0.0001365, b: 0.2097
***I*** ← Initialize individuals () ⊳$I$: initial population of individuals
$g$ = 0 ⊳count the number of generations
**while** g < max(Gen) **do** ⊳until the max generations of generations is reached
**while** each individual $i \in \boldsymbol{I}$ performs $n_{attempts}$ attempts, **do** ⊳$i$: individual in population
$i$ ← select a random individual from the population
**if** rand < $P_{SL}$**, then**
Social learning ($i$) ⊳
Access the inventory of the individual with the highest score in the population
Determine whether there exists any item in this inventory that $i$ has not found, which can be crafted using the items in $i$'s current inventory
**if** so, **then**
Randomly pick one recipe of these items, and execute
$n_{attempts} \leftarrow n_{attempts}$ - 1

**else**
IndividualModel ($i$) ⊳ Based on Algorithm 2
$n_{attempts} \leftarrow n_{attempts}$ - 1

**end if**
**end while**
$I$ ← updateModels ($I$) ⊳ Update semantic models using all successful recipes acquired via individual and social learning so far
**for each** $i \in \boldsymbol{I}$ **do**
**if** rand < $P_D$**, then**
Reinitialize ($i$) ⊳Probabilistic death event for each individual
**end if**
**end for**
$K$ ← Selection ($I$) ⊳ Selection of parents (surviving individuals) based on the probability proportional to their success
$O$ ← Reproduction (K) ⊳ Multiple copies of selected parents are initialized as their offspring
$I$ ← join (K, O) ⊳ The updated population consists of parents and their offspring.
$g$ ← g + 1
**end while**
**end function**

**Algorithm 1.** Pseudocode of the cumulative cultural evolution process.

**function IndividualModel** $(i)$
$P_S$: probability of using the semantic model
$P_G$: probability of using the generalization strategy
$M_i$: semantic model of the individual
$L_i$: inventory of the individual
$n$ = randint([1, 2, 3]) ⊳ Select randomly how many items in an innovation attempt to try n ∈ {1, 2, 3}
$T \leftarrow \{\}$ ⊳ Initialize the set of items for the innovation attempt
**if** rand > $P_S$**, then**
$T \leftarrow$ Select n items from $L_i$ with uniform probability and replacement
**else if** rand > $P_G$**, then**
$t_1 \leftarrow$ Select the first item from the inventory $L_i$ randomly with uniform probability
**if** n = 1 **then**
$T \leftarrow \{t_1\}$
**end if**
**if** $n$ > 1 **then**
$t_2 \leftarrow$ Predict $(M_i, t_1)$ ⊳Use the semantic model to predict the item that could be combined with the first item
$T \leftarrow \{t_1, t_2\}$
**if** $n$ = 3 **then**
$t_3 \leftarrow$ Predict $(M_i, t_2)$ ⊳Use the semantic model to predict the item that could be combined with the second item
$T \leftarrow \{t_1, t_2, t_3\}$
**end if**
**end if**
**else**
$T \leftarrow$ Select one successful recipe from memory uniformly at random
$t_j \leftarrow$ Select one item from T uniformly at random
$t_k \leftarrow$ Select the item that is closest to $t_j$ based on the embeddings
replace $t_j$ with $t_k$ in $T$
**end if**
**if** $T$ is not in memory, **then**
**if** $T$ leads to innovation, **then**
Add the innovation to $L_i$
**end if**
AddMemory $(T)$ ⊳ Add the attempted item combination $T$ to the memory of the individual
**end if**
**end function**

**Algorithm 2.** Pseudocode of the individual-level innovation process.

# 1 Supporting methods

## Agent-based model

**Cumulative cultural evolution models.** Pseudo-code for the CCE and individual semantic models are provided in Algorithms 1 and 2. For additional details, see the code provided with the paper.

**Semantic knowledge model.** An Artificial Neural Network (ANN) was built to learn distributed semantic representations (3–6) of the items used in the innovation task. The model is a feedforward neural network with a single hidden layer consisting of 16 neurons, which estimates the conditional probability distribution p(y|x) for a given input item x using the following transformation: $h=\sigma(W_1 x)$, $y=\phi(W_2 h)$, σ is the ReLU activation function, and Φ is the softmax function defined as $\phi(z)=\frac{e^{z_i}}{\sum_{j=1}^{n} e^{z_j}}$. The model is trained using cross-entropy loss: $L=-\Sigma t \log(z)$. Model parameters and semantic representations are optimized via back propagation defined as: $\frac{\partial L}{\partial W}=\frac{\partial L}{\partial y}x^T$, $\frac{\partial L}{\partial x}=x^T\frac{\partial L}{\partial y}$.

## Experiment

**Task interface and design.** Figure S17 shows the interface of our task, where participants discover innovations by combining existing items. Combinations are attempted in panel (A), and successful attempts produce new items added to the participant's inventory (B). In group conditions, participants can view each other's scores (C), and clicking on another participant reveals the innovations they have discovered (D). Clicking on a specific item shows its required ingredients, allowing innovations to spread via social learning (8). The interface for the non-semantic conditions was identical to that of the semantic conditions, except that the semantically meaningful images were replaced with abstract symbols (see Figure S17, right panel). Importantly, the underlying combination recipes were identical across both semantic and non-semantic conditions.

Compared to the original study (2), we made several modifications to adapt the task for online participation within a shorter duration (10 minutes). These included simplifying innovation recipes by reducing item variations (e.g., two types of axes to one), limiting combinations to a maximum of three items, removing recipes requiring repeated use of the same item, and keeping items in the inventory after use rather than removing them.

**Task structure.** The innovation task features 11 innovation levels, each building hierarchically on the previous one. It begins with initial items, followed by subsequent levels containing items derived from earlier stages. The item count per level is as follows: 6, 4, 2, 2, 2, 3, 3, 7, 11, 48, and 96. The score of finding one item is determined by the exponential function of its innovation level.

**Size of action space.** With a given inventory, participants can make innovation attempts involving individual items and combinations of two or three items. With repetition and disregarding the order, the total number of combinations for *n* items is expressed by the binomial coefficient as $\binom{n+k-1}{k}$, where $\binom{x}{y} = \frac{x!}{y!(x-y)!}$. For k = {1,2,3}, this adds up to a total of $\binom{n}{1} + \binom{n+1}{2} + \binom{n+2}{3}$. For example, for the initial inventory of 6 items, there are 83 possible unique actions. The action space expands with the increase in inventory size (total of inventory size = 184, corresponding action space = 17020).

**Random bots.** Random bots first decide how many items to combine (one, two, or three) with equal probability, then select the corresponding items uniformly at random from their inventories. In group conditions, bots also engage in success-biased social learning: they first inspect the inventory of the highest-scoring bot to determine if there are innovations to copy. If none are available, they revert to individual exploration. To ensure comparability with human performance, the number of unique attempts performed by bots was matched to that of human participants in the non-semantic condition, across both individual and group settings (averaging 68 and 57 unique attempts, respectively).

**Image similarity.** To assess the distinctiveness of the images used in the semantic and non-semantic conditions, we employed three complementary similarity measures: histogram-based (9), structural similarity (10), and encoder-based (11) (see Figure S19). In the semantic condition, many "totem" images exhibited high visual similarity, differing primarily in color and carved shapes. Accordingly, we report the distributions of image similarity both including and excluding these images (first and second rows, respectively).

Overall, images in the non-semantic conditions exhibited lower similarity compared to those in semantic conditions according to structural (mean = 0.38, sd = 0.1 vs mean = 0.46, sd = 0.25) and encoder-based (mean = 0.53, sd = 0.09 vs mean = 0.66, sd = 0.19) metrics. Conversely, histogram-based color similarity was slightly higher in non-semantic conditions (mean = 0.24, sd = 0.24 vs mean = 0.2, sd = 0.29). These results indicate that images in the non-semantic conditions are at least as discriminative as those in semantic conditions across these metrics. For subsequent behavioral strategy and representational similarity analyses, we selected structural similarity and histogram-based color distribution similarity as the primary metrics, thereby controlling for potential confounds arising from participants' preferences for combining visually dissimilar items. We excluded the encoder-based metric, as it also uses semantic properties of the images.

**Behavioral Representational Similarity Analysis (RSA).** To examine how item-level features influence participants' behaviors, we conducted representational similarity analysis inspired by approaches from neuroscience (12). For each participant, we first constructed behavioral matrices by counting the frequency with which each pair of items was combined. We then generated three model representational similarity matrices:  based on two perceptual similarities (color and structural, see Image similarity) and the other on semantic similarity (see Figure S21). In both cases, similarity was computed using the cosine similarity of vector representations of item pairs. Multiple linear regression was then used to assess whether these three model matrices significantly accounted for the variance in participants' behavioral matrices. To rule out the possibility that the model predictions are being influenced by the task structure, we masked out all successful combinations from the behavioral matrices, thereby concentrating solely on exploratory, unsuccessful attempts.

**Behavioral strategies defined in the empirical study.** In addition to perceptual and semantic similarity, we analysed the influence of a set of other strategies. Based on prior studies and our task design, we quantified the following behavioral strategies:

- Item Position. Items were displayed horizontally on the screen (see Figure S17). To capture potential positional bias, each item was coded by its grid location, with the far left assigned a value of 1.

- Number of items. Innovations could be attempted with one, two, or three items. This feature recorded the number of items used in each attempt.

- Reward. Each newly discovered item increased the participant's score. The reward feature was defined as the score associated with each item (0 for all initial items).

- Uncertainty. Participants may have preferred combining items that they were uncertain about (had not been frequently used). For each item, the uncertainty was calculated based on $\sqrt{\frac{\log(T)}{t_e + 1}}$, where $T$ denotes the total number of attempts and $t_e$ denotes the number of times the item had been selected at time $t$.

- Recency. To capture reliance on recent experience, this feature was defined as the number of attempts since an item was last selected.

- Success. This feature tracked how often an item had previously appeared in successful combinations.

- Empowerment. Item-level empowerment was calculated as the total number of successful recipes in the task structure that included the item, minus the number of such recipes already discovered. This differs from prior empowerment analyses (13), which estimated empowerment directly from the action (i.e., the combination) using neural networks. Because our task structure was relatively shallow, the data-driven approach was not feasible, as it requires large numbers of successful innovations.

For each attempt, all features were computed both for the participant's actual combination and for randomly sampled combinations from the full action space. Since the number of items used per attempt varied, feature values were averaged across all items included in the action and then normalized within participants.

# 2 Supporting results

## 2.1 ABM sensitivity and robustness analyses

We conducted extensive sensitivity and robustness analyses of the ABM parameter values and model assumptions (see Figures S2-S16 for details and results).

## 2.2 Behavioral RSA

Our main text results using multi-level logistic regression showed that participants in the semantic condition favored semantically dissimilar items, while participants in the non-semantic condition favoured perceptually dissimilar images. We therefore expect a negative contribution of the semantic similarity matrix to the behavioral matrix in the semantic condition and a negative contribution of the visual similarity matrix in the non-semantic condition. As expected, the coefficient of semantic similarity is significantly below 0 ($t(609) = -20.14$, $p < 0.001$) in the semantic condition. Conversely, the coefficient of visual structural similarity is significantly below 0 ($t(609) = -18.28$, $p < 0.001$) in the non-semantic condition. In both cases, the two coefficients differ significantly between conditions ($t(609) = -16.70$, $p < 0.001$; $t(609) = 16.83$, $p < 0.001$), in agreement with the main text analyses. Together, these analyses demonstrate that our results are robust to different analytical approaches.

## 2.3 Social learning

To confirm that the overall tendency to use social learning did not differ between the semantic and non-semantic conditions, we compared three alternative indices of social learning: the frequency of inspecting other group members' inventories, the frequency of checking the recipe of a specific item, and the frequency of immediate reproduction of the item inspected. No significant differences were found between the two conditions in any of these metrics.

Consistent with the result reported in the main text (see Figure 5), the score difference between the group maximum and participant's own scores significantly predicted an increase in the probability of engaging in social learning ($\beta = 0.02$, $SE = 0.002$, $z = 7.80$, $p < 0.001$), with a stronger effect in the non-semantic condition ($\beta = -0.01$, $SE = 0.003$, $z = -4.12$, $p < 0.001$) (see Figure S24). To investigate from whom participants chose to learn, we examined whether they preferentially inspected the inventory of the group member with the highest score. The probability of learning from the top-scoring demonstrator was significantly above chance ($\beta = 0.35$, $SE = 0.09$, $z = 3.83$, $p < 0.001$). But this tendency did not differ between the semantic and non-semantic conditions, as reflected by a non-significant interaction effect ($\beta = -0.11$, $SE = 0.12$, $z = -0.96$, $p = 0.34$).

## Supporting References